\documentclass[aps,prb,reprint,twocolumn,amsmath,amssymb,citeautoscript,longbibliography]{revtex4-2}

\usepackage{graphicx}% Include figure files
\usepackage{dcolumn}% Align table columns on decimal point
\usepackage{bm}% bold math
\usepackage{latexsym,epsfig}
\usepackage{graphicx}
\usepackage{verbatim}
\usepackage{comment}
\usepackage{amsmath}
\usepackage{amssymb}
\usepackage{physics}
\usepackage{stmaryrd}
\usepackage{color}
\usepackage{graphicx}
\usepackage{epstopdf}
\usepackage{grffile}
\usepackage{lipsum}
\usepackage{enumitem}
\usepackage{tabularx}

\usepackage[dvipsnames]{xcolor}
\DeclareGraphicsExtensions{.eps}

\newcommand{\beq}{\begin{equation}}
\newcommand{\eeq}{\end{equation}}
\newcommand{\bea}{\begin{eqnarray}}
\newcommand{\eea}{\end{eqnarray}}
\newcommand{\ben}{\begin{eqnarray*}}
\newcommand{\een}{\end{eqnarray*}}
\newcommand{\bfig}{\begin{figure}}
\newcommand{\efig}{\end{figure}}

\usepackage{hyperref}
\hypersetup{
    colorlinks=true,      
    urlcolor=blue,
    citecolor=blue,
    linkcolor=blue
}

\usepackage{booktabs}
\usepackage{array}
\usepackage{lipsum} % For dummy text, optional
\usepackage{array} % Add this to your preamble

\begin{document}
\title{ Engineering Giant Thermoelectric Performance through Electrode-Coupling
Geometry and Magnetic Flux in Quasiperiodic Su-Schrieffer-Heeger Rings} 
\author{Sridhar$^1$}\email{a22ph09005@iitbbs.ac.in}\thanks{These authors contributed equally to this work.}
\author{Souvik Roy$^1$}\email{souvikroy138@gmail.com} \thanks{These authors contributed equally to this work.}
\author{Malay Bandyopadhyay$^1$}\email{malay@iitbbs.ac.in}
\affiliation{$^1$School of Basic Sciences, Indian Institute of Technology Bhubaneswar, Odisha, India 752050}
\date{\today}

\begin{abstract}
We investigate coherent thermoelectric transport in magnetic-flux-threaded quasiperiodic Su-Schrieffer-Heeger (SSH) rings with engineered multi-site electrode couplings using the nonequilibrium Green's function formalism within the Landauer-B\"uttiker framework. We demonstrate that the electrode-coupling geometry serves as a powerful control parameter for tailoring quantum interference, thereby reshaping the transmission spectrum and thermoelectric response. In the absence of magnetic flux, the trivial dimerized phase ($t_1>t_2$) exhibits the highest thermoelectric efficiency, with asymmetric coupling producing a substantially larger figure of merit than the symmetric geometry. Magnetic flux further reconstructs the transmission spectrum through Aharonov-Bohm interference, driving a crossover of the optimal thermoelectric regime from the trivial to the topological dimerized phase. Under optimal flux conditions, the thermoelectric figure of merit reaches $ZT \approx 12$ for symmetric coupling and is dramatically enhanced to $ZT \approx 90$ for asymmetric coupling through enhanced energy filtering and suppressed electronic thermal transport. We further establish a clear correlation between the enhancement of thermoelectric efficiency and the violation of the Wiedemann-Franz law. Our results demonstrate that the combined interplay of quasiperiodicity, topology, magnetic flux, and electrode-coupling geometry provides a versatile strategy for engineering high-performance coherent thermoelectric devices.

\end{abstract}

\maketitle
\section{Introduction}

Thermoelectric devices offer a promising route for directly converting waste heat
into electrical energy and have attracted considerable interest as sustainable energy-harvesting technologies~\cite{Benenti2017, PhysRevLett.112.130601, PhysRevB.91.115425, Mahan1996, Dresselhaus2007, Hicks1993, Bedkihal2025,PhysRevB.83.085428,Sothmann_2015,PhysRevLett.106.230602}. Their performance is governed by the competition between electrical and
thermal transport, quantified by the dimensionless figure of merit, where $S$, $G$, and $\kappa$ denote the Seebeck coefficient, electrical conductance, and thermal conductance, respectively. Enhancing $ZT$ therefore requires simultaneously promoting energy-selective charge transport and suppressing unwanted heat transport ~\cite{Bergfield2010, Karlstrom2011, Trocha2012, Whitney2014}. Within the Landauer-Buttiker framework, thermoelectric coefficients are governed by the energy dependence of the electronic transmission function.~\cite{Dhar2006, Datta1995, Datta2005}. Quantum interference provides a powerful route toward this goal by reshaping the energy dependence of the electronic transmission. In coherent mesoscopic systems, sharp resonances, antiresonances, and asymmetric transmission profiles can enhance thermopower while suppressing thermal transport, enabling substantial improvements in thermoelectric performance. Such interference-induced energy filtering has been explored in quantum dots \cite{Karlstrom2011, Miroshnichenko2010}, molecular junctions \cite{Miao2018, Perroni2016, Lambert2015}, mesoscopic rings \cite{GarciaSuarez2013}, and topological nanostructures~\cite{Karlstrom2011, Trocha2012}.

Among low-dimensional systems, the Su-Schrieffer-Heeger (SSH) model \cite{PhysRevLett.42.1698, PhysRevB.22.2099, RevModPhys.60.781} provides a particularly useful platform for controlling coherent transport through the interplay of dimerization and topology~\cite{PhysRevLett.42.1698, Asboth2016}. The relative strengths of the intracell and intercell hoppings define distinct trivial and topological regimes with different spectral and transport characteristics~\cite{PhysRevA.80.021603, Harper1955}. Introducing quasiperiodicity further enriches the transmission spectrum through band fragmentation, localization, and resonant conducting states, offering additional possibilities for energy-selective transport~\cite{svkprb2} In ring geometries, magnetic flux introduces an Aharonov-Bohm phase between different propagation paths, enabling continuous control of constructive and destructive quantum interference and, consequently, of the transmission spectrum and
thermoelectric response.~\cite{Aharonov1959, Buttiker1984, Roy2025FluxDriven, Roy2023CircularCurrent,svkprb1}. The combined interplay of dimerization, quasiperiodicity, and flux therefore provides a natural setting for engineering coherent thermoelectric transport~\cite{Aharonov1959, Buttiker1984, Bhattacharya2026Thermoelectric}.

In addition to the intrinsic properties of the lattice, the geometry of the system–electrode coupling provides an independent means of controlling coherent transport. Conventional transport studies typically employ single-site contacts \cite{Chakraborty2024Thermoelectric, Ganguly2024Thermoelectric}, whereas coupling an electrode to multiple sites can create additional interference pathways and substantially reshape the transmission spectrum \cite{Sridhar2026Interplay, l9gd-k9yw}. The spatial arrangement of the contacts, therefore, acts as an
external control over the interference pattern, allowing the coupling between the system's internal structure and the reservoirs to be tailored. Despite its potential, the role of electrode-coupling geometry in thermoelectric transport has received considerably less attention, particularly in systems where topology, quasiperiodicity, and magnetic-flux-induced interference
coexist.

This motivates a systematic investigation of how electrode-coupling geometry can be combined with the intrinsic control parameters of quasiperiodic topological systems to optimize thermoelectric transport. Quantum confinement, energy filtering, and quantum interference have already been shown to enhance thermoelectric performance in nanoscale systems, with sharp transmission features in quantum dots, molecular junctions, topological systems, and quasiperiodic lattices providing favorable conditions for thermoelectric conversion.~\cite{Bhandari2025Multidot, Karlstrom2011, Trocha2012, Whitney2014, Verma2022NonEquilibrium}. Nevertheless, the thermoelectric figure of merit in coherent mesoscopic systems typically
remains in the range $ZT \sim 1-30$. \cite{Sitek2013Dicke, l9gd-k9yw, Bhandari2025Multidot, Bergfield2009Thermoelectric, Bennett2024Quantum, Bhattacharya2026Thermoelectric, Xu2014, Zhou2025Interaction} depending on the device architecture and transport regime. This raises the possibility that engineering the spatial distribution of electrode contacts, together with magnetic-flux-induced interference and the system's intrinsic topology and quasiperiodicity, could provide an additional degree of control over energy-selective transport. In particular, such a combined strategy may enable tailoring of the transmission spectrum to enhance thermopower while suppressing thermal transport, potentially accessing thermoelectric regimes beyond those achievable through intrinsic lattice engineering alone.

In this work, we investigate coherent thermoelectric transport in a magnetic-flux-threaded quasiperiodic Su-Schrieffer-Heeger (SSH) ring with engineered multi-site electrode-coupling geometries. We consider symmetric coupling, in which three lattice sites are connected to each reservoir, and asymmetric coupling, in which three sites are connected to the source and a single site to the drain. Using the nonequilibrium Green’s function formalism within the Landauer-Buttiker framework~\cite {Landauer1957, Datta1995, Datta2005, Buttiker1986, Meir1992, Haug2008, Fisher1981, Behera2023Quantum, Bedkihal2013FluxDependent}, we systematically examine the transmission spectrum and the resulting electrical conductance, Seebeck coefficient, electronic thermal conductance, thermoelectric figure of merit, and Lorenz ratio. We demonstrate that the electrode-coupling geometry directly controls quantum interference and strongly influences the optimal
thermoelectric regime. In the absence of magnetic flux, the optimal response is found in the trivial dimerized phase. In contrast, magnetic flux shifts the optimal regime toward the topological phase through Aharonov-Bohm interference. This effect is dramatically amplified by asymmetric electrode coupling, yielding $ZT_{\max} \approx 90$ through enhanced energy filtering and strongly suppressed electronic thermal conductance. We further show that the regions of enhanced thermoelectric performance coincide with strong violations of the Wiedemann-Franz law, highlighting the decoupling of charge and heat transport. These results establish electrode-coupling geometry and magnetic flux as complementary control parameters for engineering
coherent thermoelectric transport in quasiperiodic topological systems.

The remainder of this paper is organized as follows. Section~\ref{secII} introduces the quasiperiodic SSH ring model and outlines the nonequilibrium Green's function formalism employed to determine the transmission probability and the corresponding thermoelectric coefficients, including the electrical conductance, Seebeck coefficient, electronic thermal conductance, thermoelectric figure of merit, and Lorenz ratio. In Sec .~\ref {secIII}, we investigate the thermoelectric transport
properties for various system-electrode coupling geometries, demonstrating how coupling geometry governs the transport behavior. Finally, Sec .~\ref {secIV} presents the concluding remarks.

\section{Model and Hamiltonian}
\label{secII}
We consider a finite quasiperiodic Su-Schrieffer-Heeger (SSH) ring coupled to two noninteracting metallic reservoirs, denoted as the source (S) and drain (D), as schematically illustrated in Fig.~\ref{fig1}. The ring combines SSH dimerization with an Aubry-Andr\'e-Harper (AAH) quasiperiodic onsite potential and is threaded by a magnetic flux $\phi$. The coupling between the ring and the reservoirs is engineered through different multi-site contact geometries, allowing us to investigate the interplay
of lattice structure, quasiperiodicity, Aharonov-Bohm interference, and electrode-coupling geometry in coherent thermoelectric transport. The total number of lattice sites is denoted by $N$, with $N=2L$, where $L$ is the number of SSH unit cells. The total Hamiltonian is written as

\begin{equation}
\mathcal{H}
=
\mathcal{H}_{R}
+
\mathcal{H}_{B}
+
\mathcal{H}_{\mathrm{int}},
\end{equation}
where $\mathcal{H}_{R}$ describes the quasiperiodic SSH ring, $\mathcal{H}_{B}$ represents the source and drain reservoirs, and $\mathcal{H}_{\mathrm{int}}$ describe the coupling between the ring and the reservoirs.

\subsection{Quasiperiodic SSH Ring}

The Hamiltonian of the SSH ring within the nearest-neighbor tight-binding approximation is

\begin{align}
\mathcal{H}_{R}
=&
\sum_{i=1}^{N}
\epsilon_i c_i^\dagger c_i +\sum_{i=1}^{N/2}\Big(t_1 e^{i\phi_i}c_{2i-1}^{\dagger}c_{2i}
+
t_2 e^{i\phi_i}
c_{2i}^{\dagger}c_{2i+1}
\nonumber\\
&+\mathrm{H.c.}
\Big).
\label{eq:HR}
\end{align}

Here, $c_i^\dagger$ ($c_i$) creates (annihilates) an electron at lattice site (i). The hopping amplitudes $t_1$ and $t_2$ correspond to the intracell and intercell hoppings, respectively, defining the dimerized SSH lattice. Periodic boundary conditions are imposed, $c_{N+1} \equiv c_1$, thereby forming a closed ring.
The dimerization parameter distinguishes the topological ($t_1 < t_2$), uniform ($t_1 = t_2$), and trivial ($t_1 > t_2$) regimes.

The quasiperiodicity is introduced through an Aubry-Andr\'e-Harper (AAH) modulation of the onsite energies,

\begin{equation}
\epsilon_i
=
W \cos \left( 2\pi b i + \phi_{\mathrm{AAH}} \right),
\end{equation}

where $W$ is the modulation amplitude, $b=(\sqrt{5}-1)/2$ is the irrational modulation parameter corresponding to the golden mean, and $\phi_{\mathrm{AAH}}$ is the phase of the quasiperiodic potential, which is fixed at zero throughout this work. The irrational modulation generates a deterministic quasiperiodic potential and introduces correlated disorder into the system, thus modifying the spectral and spatial structure of the electronic states.

A magnetic flux $\Phi$ threading the ring introduces an Aharonov-Bohm (AB) phase into the hopping amplitudes. The accumulated phase is given by

\begin{equation}
\sum_{i=1}^{N}
\phi_{i,i+1}
=
2\pi\frac{\phi}{\phi_0},
\end{equation}
where $\phi_0=\frac{h}{e}$ is the magnetic flux quantum. The individual bond phases depend on the choice of gauge, whereas their sum around the ring
determines the Aharonov-Bohm phase acquired by an electron traversing the closed loop. The magnetic flux, therefore, provides a tunable control of quantum interference between different propagation pathways. For the numerical calculations, we use a finite ring with $N=34$ sites, which provides a convenient
finite approximation to the irrational golden-mean modulation. This system size is sufficiently large to capture the quasiperiodic behavior while keeping the numerical calculations tractable.
\subsection{Reservoirs and System-Electrode Coupling}
The source (S) and drain (D) reservoirs are modeled as noninteracting electron baths,

\begin{equation}
\mathcal{H}_{B}
=
\sum_{\alpha=S,D}
\sum_{k}
\epsilon_{k\alpha}\,
c_{k\alpha}^{\dagger}
c_{k\alpha},
\end{equation}

where $c_{k\alpha}^{\dagger}$ ($c_{k\alpha}$) creates (annihilates) an electron with momentum $k$ and energy $\epsilon_{k\alpha}$ in reservoir $\alpha$, where $\alpha=S,D$ denote the source and drain reservoirs, respectively.

The coupling between the SSH ring and the reservoirs is described by

\begin{equation}
\mathcal{H}_{\mathrm{int}}
=
\sum_{\alpha=S,D}
\sum_{k}
\left(
v_{ik}^{\alpha}
c_i^{\dagger}
c_{k\alpha}
+
\mathrm{H.c.}
\right),
\end{equation}

where $v_{ik}^{\alpha}$ denotes the tunneling amplitude between lattice site $i$ of the SSH ring and the reservoir state $(k,\alpha)$. The set of coupled lattice sites, $i\in\alpha$, is determined by the electrode-coupling geometry. We consider two different coupling geometries. In the symmetric coupling geometry, three
lattice sites are coupled to each reservoir. The source is connected to sites
$(1,2,N)$, while the drain is connected to sites $(N/2-1,N/2+1,N/2+2)$. In the asymmetric coupling geometry, the source remains coupled to the three sites $(1,2, N)$, whereas the drain is coupled only to the single site $(N/2+1)$. Thus, the two geometries provide different sets of coherent propagation pathways between the source and drain and consequently generate distinct interference patterns in the
transmission spectrum.
\subsection{Thermoelectric Transport}
We employ the nonequilibrium Green's function (NEGF) formalism \cite{Dhar2006, Landauer1957, Datta1995, Datta2005, Buttiker1986, Meir1992, Haug2008, Fisher1981, Behera2023Quantum, Bedkihal2013FluxDependent} to describe coherent transport through the ring. The retarded Green's function of the device is \cite{l9gd-k9yw, Sridhar2026Interplay}. 

% \begin{equation}
% \begin{aligned}
% G^{\pm}(E)
% =
% \Bigg[
% EI-\mathcal{H}_R
% &-\sum_{m^\prime}
% \Sigma_{m^\prime,m}^{\pm(S)}(E)\,\delta_{mi}
% \\
% &-\sum_{n^\prime}
% \Sigma_{n^\prime,n}^{\pm(D)}(E)\,\delta_{ni}
% \Bigg]^{-1}.
% \end{aligned}
% \end{equation}

\begin{equation}
\begin{aligned}
G^{\pm}(E)
=
\Bigg[
EI-\mathcal{H}_R
&-
\Sigma_{\alpha}^{\pm(S)}(E)\,
-
\Sigma_{\alpha}^{\pm(D)}(E)\
\Bigg]^{-1}.
\end{aligned}
\end{equation}
Where $\Sigma_{\alpha}^{+}$ is the retarded self-energy associated with reservoir $\alpha$, the advanced Green's function of the device is $G^{-}$, and $I$ is the identity matrix. Throughout this work, we employ the wide-band limit (WBL), in which the electrode density of states is assumed to be energy independent over the relevant transport window. Within this approximation, the real part of the self-energy vanishes, and the self-energy matrices become purely imaginary. The retarded self-energies are therefore given by
$\Sigma_{\alpha}^{+}=-\frac{i}{2}\Gamma_{\alpha}$, where $\Gamma_{\alpha}$ is the hybridization matrix describing the coupling between the ring and reservoir $\alpha$.
The matrix structure of $\Gamma_{\alpha}$ contains the information about both the strength and spatial distribution of the system-electrode coupling.

The transmission probability for electrons traversing the system from the source to the drain is given by the Fisher-Lee relation \cite{Fisher1981},
\begin{equation}
T_{SD}(E,\phi)=\mathrm{Tr}
\left[
\Gamma_S G^{+}(E,\phi)\Gamma_D G^{-}(E,\phi)
\right],
\end{equation}
where $\Gamma_S$ and $\Gamma_D$ are the source and drain hybridization matrices, respectively, and $G^{+}(E,\phi)$ and $G^{-}(E,\phi)$ are the corresponding retarded and advanced Green's functions of the ring.

The transmission function is the central quantity from which all transport coefficients are evaluated. In the following, we investigate the charge, heat, and thermoelectric transport properties of the quasiperiodic SSH ring for different source-drain coupling geometries. Unless otherwise specified, symmetric coupling strengths are assumed throughout, i.e., $\gamma_S=\gamma_D=\gamma$. Within the nonequilibrium Green's function (NEGF) formalism, the charge and heat currents flowing from the source ($S$) to the drain ($D$) are given by \cite{Yamamoto2015, Sivan1986, Butcher1990, Bedkihal2013, Landauer1957, Datta1995, Datta2005, Buttiker1986, Meir1992, Haug2008, Fisher1981}
\begin{equation}
I_C=
\frac{e}{h}
\int dE\,
[f_S(E)-f_D(E)]\,T_{SD}(E),
\end{equation}
\begin{equation}
I_Q=
\frac{1}{h}
\int dE\,
(E-\mu)
[f_S(E)-f_D(E)]\,T_{SD}(E).
\end{equation}
Here, $f_{\nu}(E)=[\exp((E-\mu_{\nu})/k_BT_{\nu})+1]^{-1}$ is the Fermi-Dirac distribution function, where $T_{\nu}$ and $\mu_{\nu}$ denote the temperature and chemical potential of reservoir $\nu$, respectively, with $\nu\in{S,D}$.

In the linear-response regime, the voltage and temperature biases are assumed sufficiently small compared with the average chemical potential and temperature, respectively. Specifically, we consider $\Delta\mu=\mu_S-\mu_D$ and $\Delta T=T_S-T_D$ satisfying $|\Delta\mu|/T \ll 1$ and $|\Delta T|/T \ll 1$, where $\Delta V=\Delta\mu/e$ is the applied voltage bias. Under these conditions, the charge and heat currents can be expressed as linear functions of the corresponding thermodynamic affinities.

\begin{equation}
\begin{pmatrix}
I_C \\[4pt]
I_Q
\end{pmatrix}
=
\begin{pmatrix}
e^2I_0 & eI_1 \\[4pt]
eI_1 & I_2
\end{pmatrix}
\begin{pmatrix}
\Delta V \\[4pt]
\dfrac{\Delta T}{T}
\end{pmatrix}
\end{equation}
where:
\begin{equation}
I_n=\frac{1}{h}\int_{-\infty}^{\infty} dE
(E-\mu)^n
T_{SD}(E,\phi)
\left(-\frac{\partial f}{\partial E}\right),
\end{equation}
where $-\frac{\partial f}{\partial E}=\frac{1}{4k_BT}
\operatorname{sech}^2
\left(
\frac{E-\mu}{2k_BT}
\right)$
 is the derivative of the Fermi-Dirac distribution function. In terms of these transport integrals, the electrical conductance and electronic thermal conductance are given by $G=e^2 I_0$ and $k_e=
\frac{1}{T}
\left(
I_2-\frac{I_1^2}{I_0}
\right)$,
respectively. The remaining thermoelectric coefficients can also be expressed in terms of the Onsager integrals. In particular, the Seebeck coefficient (thermopower) is given by $S=\frac{I_1}{ eT I_0}$, while the thermoelectric figure of merit is defined as
\begin{equation}
ZT=
\frac{I_1^2}
{I_0I_2-I_1^2}.
\end{equation}

To characterize the departure from conventional charge-heat transport relations, we consider the Lorenz number $L$. For a conventional free-electron system, the electrical and thermal conductances are related by the Wiedemann-Franz (WF) law, yielding the universal Lorenz number $L_{0}=\pi^{2}/3e^{2}$. We therefore define the normalized Lorenz ratio $R=L/L_{0}$, so that $R=1$ corresponds to the Wiedemann-Franz value, while deviations from unity quantify the violation of the Wiedemann-Franz law. Within the linear-response regime, $R$ can be expressed in terms of the Onsager integrals as follows:

\begin{equation}
R=
\frac{3}{(\pi T)^2}
\left[
\frac{I_2}{I_0}-
\left(\frac{I_1}{I_0}\right)^2
\right],
\end{equation}

The transmission function $T_{SD}(E,\Phi)$, modified by the combined effects of SSH dimerization, quasiperiodicity, magnetic flux, and electrode-coupling geometry, therefore provides the common microscopic basis for all the analyses of different thermoelectric quantities considered in the next section.

\section{Results and discussion}

\begin{figure}
    \centering
    \includegraphics[scale=0.35]{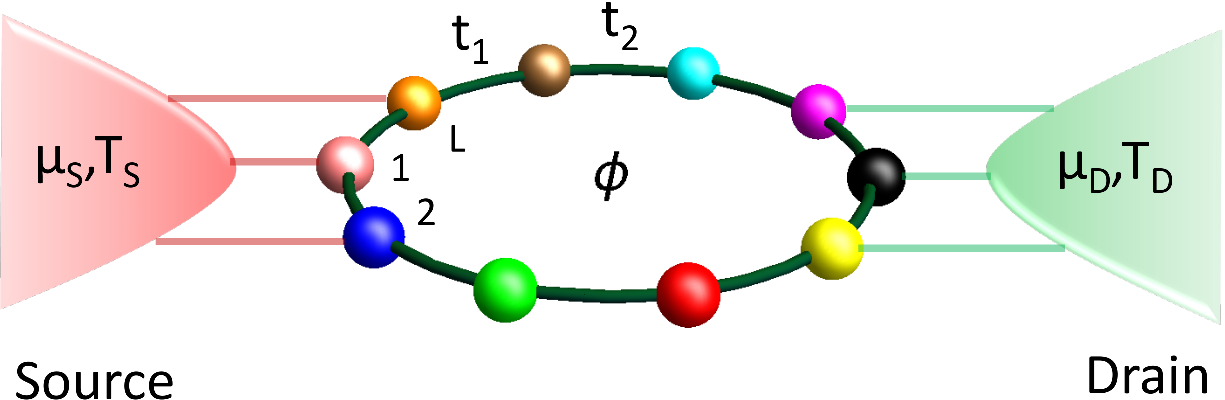}
    \caption{Schematic of the symmetric electrode-coupling geometry with three lattice sites coupled to each reservoir.}
    \label{fig1}
\end{figure}

How strongly can thermoelectric transport be controlled without changing the underlying lattice? To address this question, we first isolate the role of the system-electrode coupling geometry in the quasiperiodic SSH ring. We compare two configurations that share the same ring Hamiltonian and reservoir parameters but differ in the spatial arrangement of the contacts: a symmetric geometry with three sites coupled to each reservoir (see Fig.~\ref{fig1}) and an asymmetric geometry with three source contacts and a single drain contact (see Fig.~\ref{fig5}). This comparison provides a direct measure of how coherent pathways introduced solely by the electrode geometry reshape the transmission
spectrum and, consequently, the thermoelectric response. We begin with zero magnetic flux, allowing the effects of dimerization and quasiperiodicity to be identified independently of Aharonov-Bohm interference. We then introduce magnetic flux to explore how the resulting interference can further reorganize the energy-filtering characteristics. Throughout, we characterize the transport through the transmission function $T(E)$, electrical conductance $G$, Seebeck coefficient $S$, electronic thermal conductance $\kappa_{e}$, thermoelectric figure of merit $ZT$, and normalized Lorenz ratio $R$. Unless otherwise specified, the calculations are performed for a ring of $N=34$ sites, with the remaining parameters stated in the corresponding figures and discussions.

\label{secIII}
% \subsection{Electrode-coupling engineering of thermoelectric transport}
\subsection{Symmetric electrode-coupling geometry: Zero magnetic flux}

\begin{figure}[t]
    \centering
    \includegraphics[width=8cm,height=8cm]{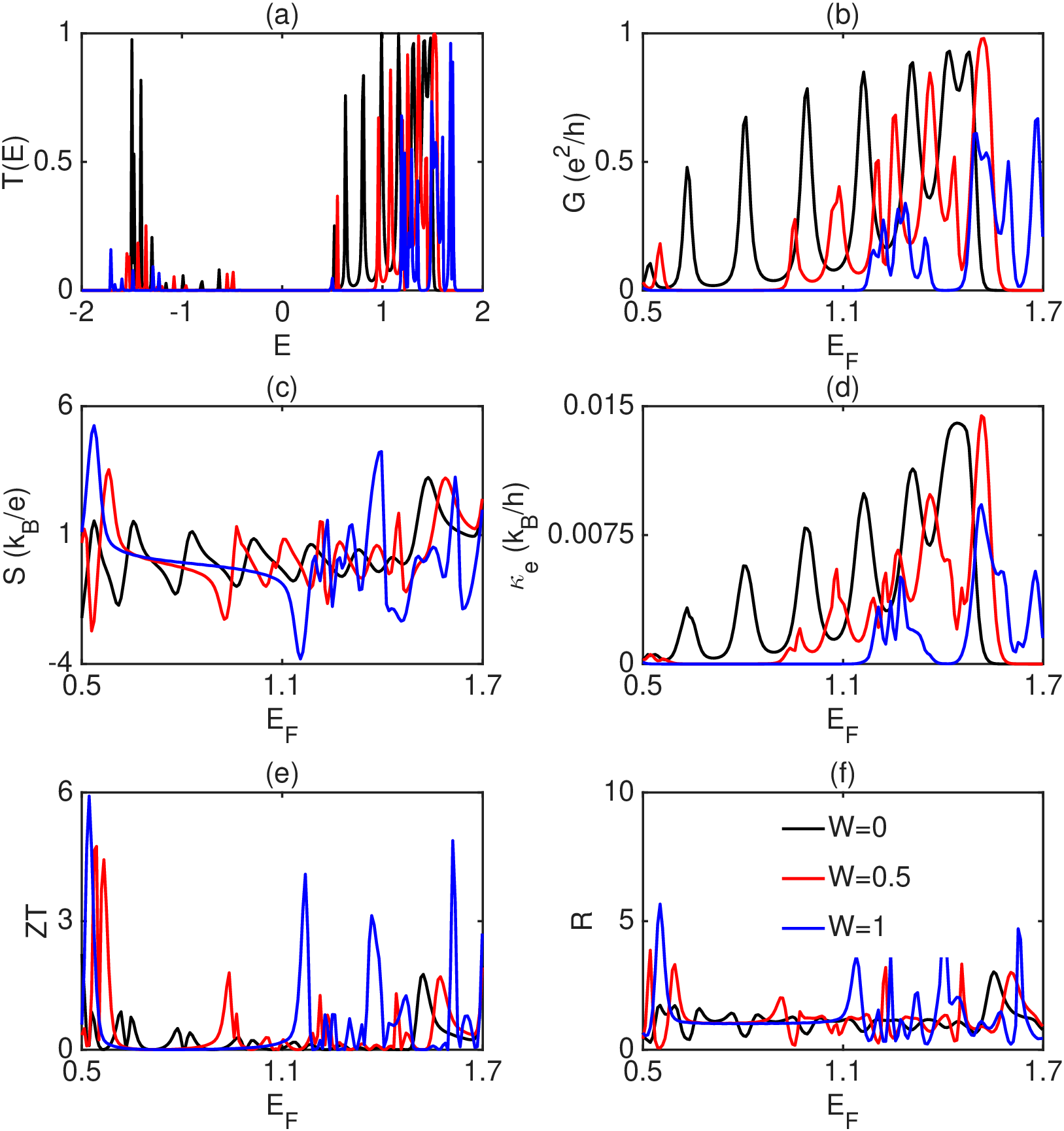}
    
    \caption{Symmetric electrode-coupling geometry in the topological dimerized regime ($t_1<t_2$). (a) Transmission spectrum, (b) electrical conductance, (c) Seebeck coefficient, (d) electronic thermal conductance, (e) thermoelectric figure of merit ($ZT$), and (f) Lorenz ratio as functions of the Fermi energy for different quasiperiodic modulation strengths. The parameters are $\gamma=0.05$, $T=0.005$, $\phi=0$, $t_1=0.5$, and $t_2=1$.}
    \label{fig2}
    
\end{figure}

\begin{figure}[t]
    \centering
    \includegraphics[width=8cm,height=8cm]{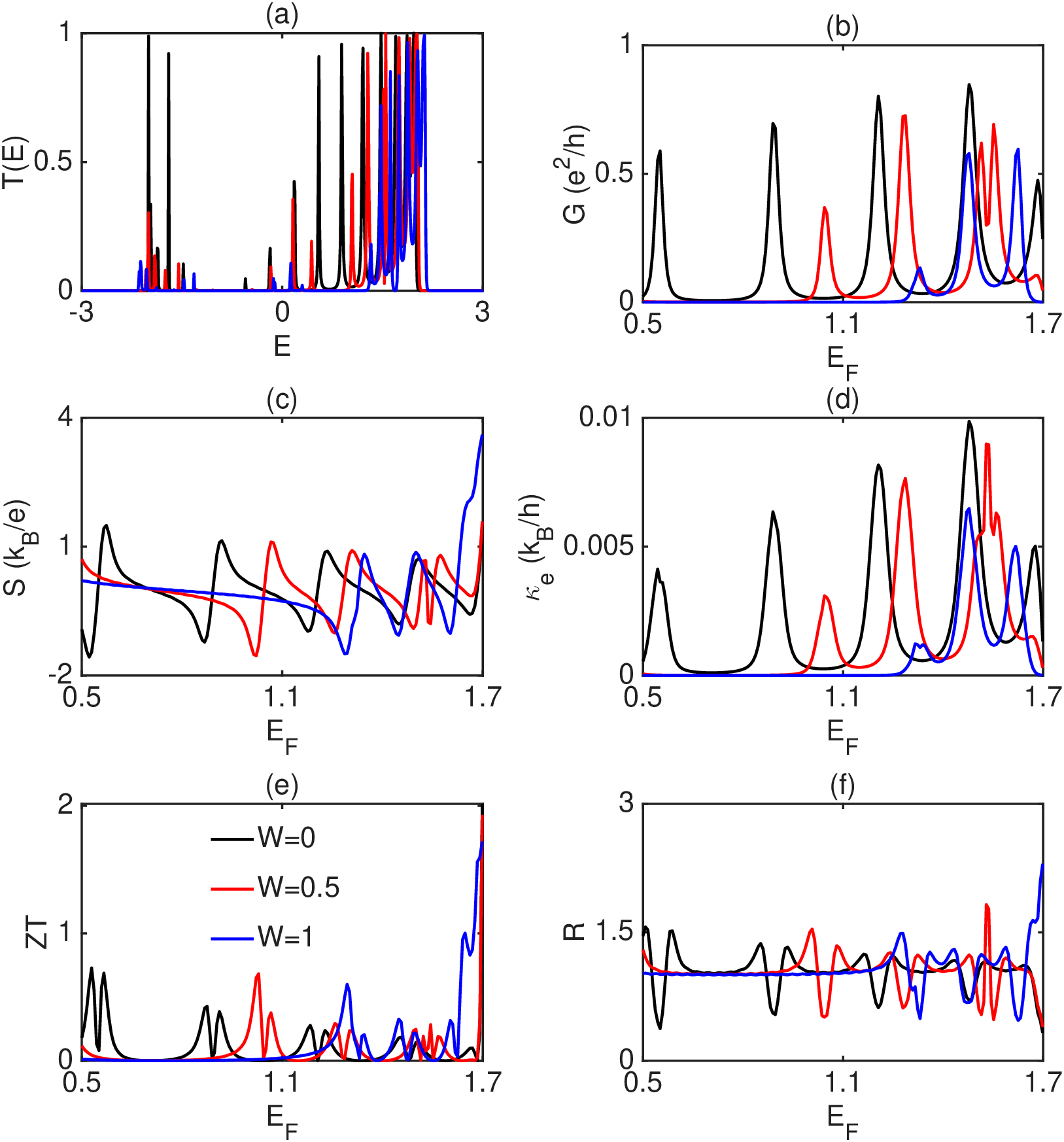}
    \caption{Symmetric electrode-coupling geometry in the homogeneous-hopping regime ($t_1=t_2$). (a) Transmission spectrum, (b) electrical conductance, (c) Seebeck coefficient, (d) electronic thermal conductance, (e) thermoelectric figure of merit ($ZT$), and (f) Lorenz ratio as functions of the Fermi energy for different quasiperiodic modulation strengths. The parameters are $\gamma=0.05$, $T=0.005$, $\phi=0$, and $t_1=t_2=1$.}
    %\caption{Ring(3,3) $t_1$ eq $t_2$}
    \label{fig3}
\end{figure}

\begin{figure}[t]
    \centering
    \includegraphics[width=8cm,height=8cm]{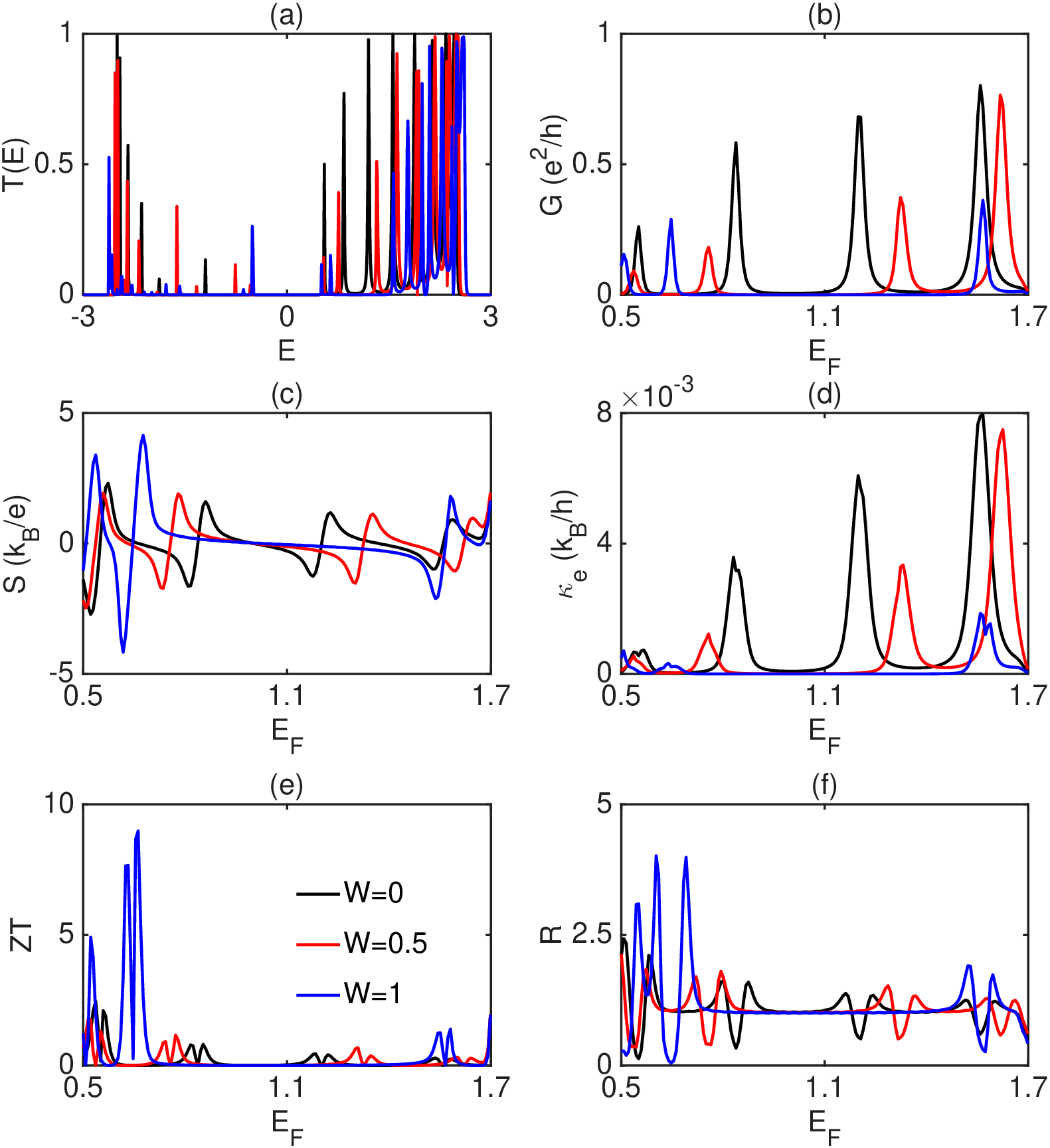}
    \caption{Symmetric electrode-coupling geometry in the trivial dimerized regime ($t_1>t_2$). (a) Transmission spectrum, (b) electrical conductance, (c) Seebeck coefficient, (d) electronic thermal conductance, (e) thermoelectric figure of merit ($ZT$), and (f) Lorenz ratio as functions of the Fermi energy for different quasiperiodic modulation strengths. The parameters are $\gamma=0.05$, $T=0.005$, $\phi=0$, $t_1=1.5$, and $t_2=1$.}
    %\caption{Ring(3,3) $t_1$ gr $t_2$}
    \label{fig4}
\end{figure}

\begin{figure}
    \centering
    \includegraphics[scale=0.35]{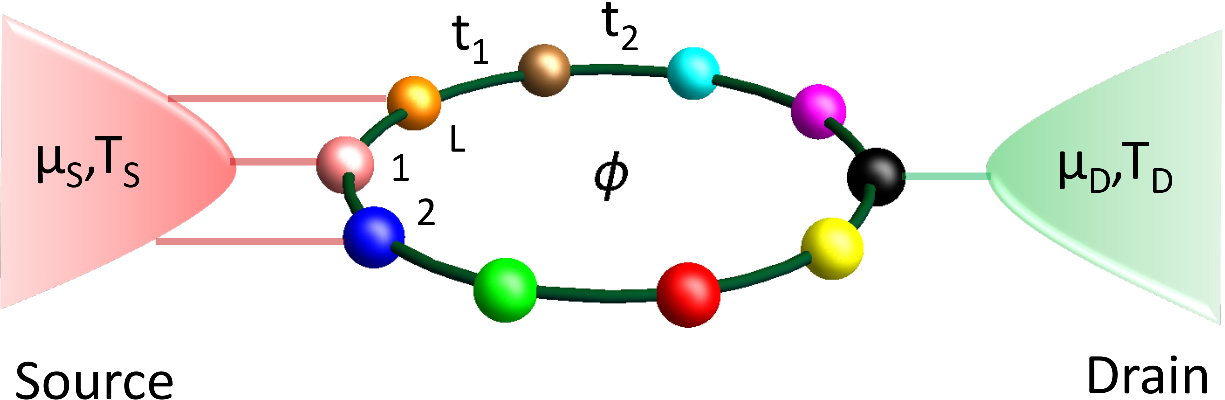
    }
    \caption{Schematic of the asymmetric electrode-coupling geometry with three lattice sites coupled to the source reservoir and a single lattice site coupled to the drain reservoir.}
    \label{fig5}
\end{figure}

We begin by focusing on the symmetric electrode-coupling geometry, in which three lattice sites are connected to both the source and drain reservoirs. The transport characteristics in the topological ($t_1 < t_2$), homogeneous ($t_1 = t_2$), and trivial dimerized ($t_1 > t_2$) regimes are depicted in Figs.~\ref{fig2}-\ref{fig4} for a range of quasiperiodic modulation strengths $W$. Here, we fix the magnetic flux at zero ($\phi = 0$) to isolate the roles of dimerization, quasiperiodicity, and
multi-site electrode coupling, free from the influence of Aharonov-Bohm interference. The presence of multiple contacts to each reservoir creates an array of coherent transport pathways through the ring, rendering the transmission spectrum highly sensitive to both the hopping configuration and the degree of quasiperiodic modulation.

% \subsubsection{Symmetric coupling geometry}

In the topological regime ($t_1 < t_2$), the transmission spectrum is notably suppressed near the band center, punctuated by several conducting resonances farther from the center [Fig.~\ref{fig2}(a)]. As the strength of the quasiperiodic modulation increases from $W = 0$, this spectrum becomes increasingly fragmented, with resonances growing both denser and more asymmetric in their distribution. In the homogeneous regime ($t_1 = t_2$), transmission is primarily confined to the central
energy region [Fig.~\ref{fig3}(a)], while in the trivial dimerized regime ($t_1 > t_2$), transmission channels emerge on either side of the spectrum [Fig.~\ref{fig4}(a)]. Thus, both dimerization and quasiperiodicity offer powerful means to directly tune the accessibility and positioning of transport channels.


The electrical conductance closely follows these changes in the transmission spectrum. In the topological regime, conductance exhibits pronounced sensitivity to the Fermi energy, with moderate quasiperiodic modulation amplifying conductance before stronger modulation ultimately suppresses it. By contrast, in both the homogeneous and trivial dimerized regimes, conductance reaches its maximum in the ordered system and diminishes steadily as $W$ increases. Ultimately, the impact of quasiperiodicity on coherent transport is intricately shaped by the interplay between dimerization and the system's spectral landscape. The thermoelectric response is dictated not only by the absolute magnitude of the transmission function, but crucially by its intricate energy dependence within the thermal window set by
the Fermi energy. As the system transitions from the topological to the trivial dimerized regime [Figs.~\ref{fig2}(d), \ref{fig3}(d), and \ref{fig4}(d)], electronic thermal conductance systematically diminishes, signaling progressively constrained pathways for heat transport. Simultaneously, the Seebeck coefficient is markedly amplified as the transmission acquires a stronger energy asymmetry. In both dimerized
regimes, this asymmetry emerges sharply at selected Fermi energies, whereas in the
homogeneous regime, it is predominantly shifted toward positive energies. Escalating the quasiperiodic modulation $W$ further accentuates the energy selectivity of the transmission, thereby amplifying thermopower and conferring a more sharply defined thermoelectric response.

These intricate interplays manifest strikingly in the thermoelectric figure of merit, $ZT$. In the topological dimerized regime, a peak $ZT$ of around 6 emerges, attributable to the synergistic enhancement of the Seebeck coefficient and the concurrent suppression of electronic thermal conductance. The trivial dimerized regime exhibits even more remarkable performance, achieving $ZT$ values approaching 9, where robust thermopower is paired with a pronounced reduction in $\kappa_{e}$. Conversely, the homogeneous regime attains a considerably lower maximum $ZT$
near 2, as its modest thermopower constrains thermoelectric efficiency, despite relatively high electrical conductance. These findings underscore a central principle: optimal thermoelectric performance is realized when thermopower is maximized and electronic thermal conductivity $\kappa_{e}$ is effectively minimized.

The normalized Lorenz ratio, $R$, serves as a sensitive probe of charge-heat transport decoupling within these systems. Across all three hopping regimes, pronounced violations of the Wiedemann-Franz law emerge, with the extent of this deviation intricately governed by the degree of dimerization. The trivial dimerized regime exhibits the most significant suppression of $R$, coinciding with the highest thermoelectric figure of merit, $ZT$. In contrast, the homogeneous regime displays the weakest Lorenz-ratio violation and correspondingly low thermoelectric efficiency, while the topological regime displays intermediate characteristics, marked by a
substantial reduction in $R$ accompanying $ZT \simeq 6$. These observations compellingly underscore the fundamental connection between Lorenz-ratio suppression and enhanced thermoelectric performance.

The remarkable alignment between Lorenz-ratio suppression and the elevation of $ZT$ unveils a fundamental physical principle governing thermoelectric behavior. Here, the intricate energy dependence of the transmission function, sculpted by the interplay of dimerization and quasiperiodicity, enables a pronounced distinction between charge and heat transport: electrical conductivity remains robust, even as thermal conductivity is selectively diminished. Strikingly, this decoupling persists even without magnetic flux, as the symmetric multi-site coupling and the quasiperiodic SSH spectrum conspire to maximize thermoelectric efficiency. Among all regimes, the dimerized configuration stands out as the most propitious, embodying the ideal balance for heightened thermoelectric performance. Ultimately, these findings reveal
that the decoupling of charge and heat transport constitutes the central mechanism behind the observed thermoelectric enhancement.

% \subsubsection{Asymmetric coupling geometry}
\subsection{Asymmetric electrode-coupling geometry: Zero magnetic flux}

\begin{figure}[t]
    \centering
    \includegraphics[width=8cm,height=8cm]{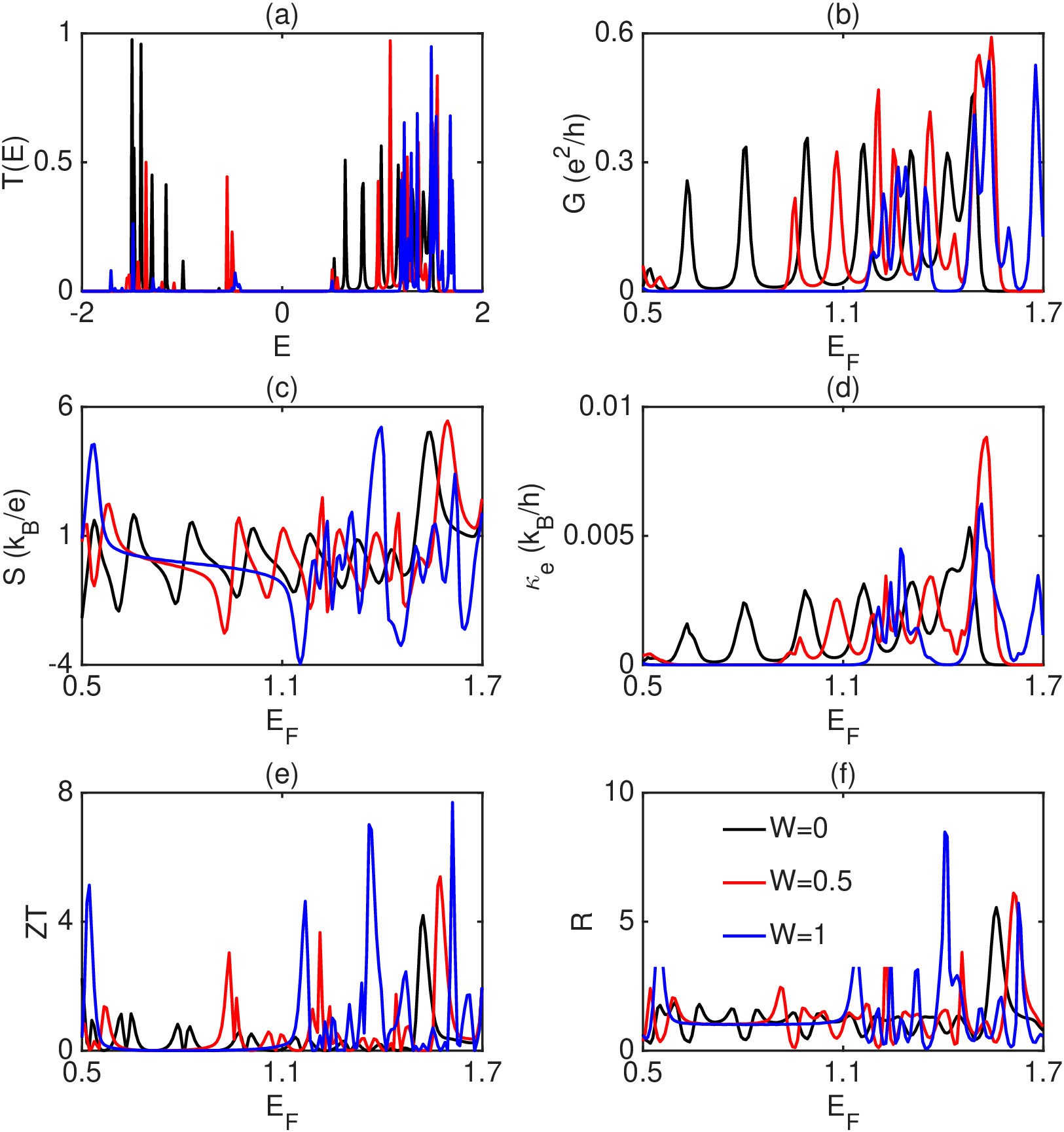}
    \caption{Asymmetric $(3-1)$ electrode-coupling geometry in the topological dimerized regime ($t_1<t_2$). (a) Transmission spectrum, (b) electrical conductance, (c) Seebeck coefficient, (d) electronic thermal conductance, (e) thermoelectric figure of merit ($ZT$), and (f) Lorenz ratio as functions of the Fermi energy for different quasiperiodic modulation strengths. The parameters are $\gamma=0.05$, $T=0.005$, $\phi=0$, $t_1=0.5$, and $t_2=1$.}
   % \caption{Ring(3,1) $t_1$ less $t_2$}
    \label{fig6}
\end{figure}

\begin{figure}[t]
    \centering
    \includegraphics[width=8cm,height=8cm]{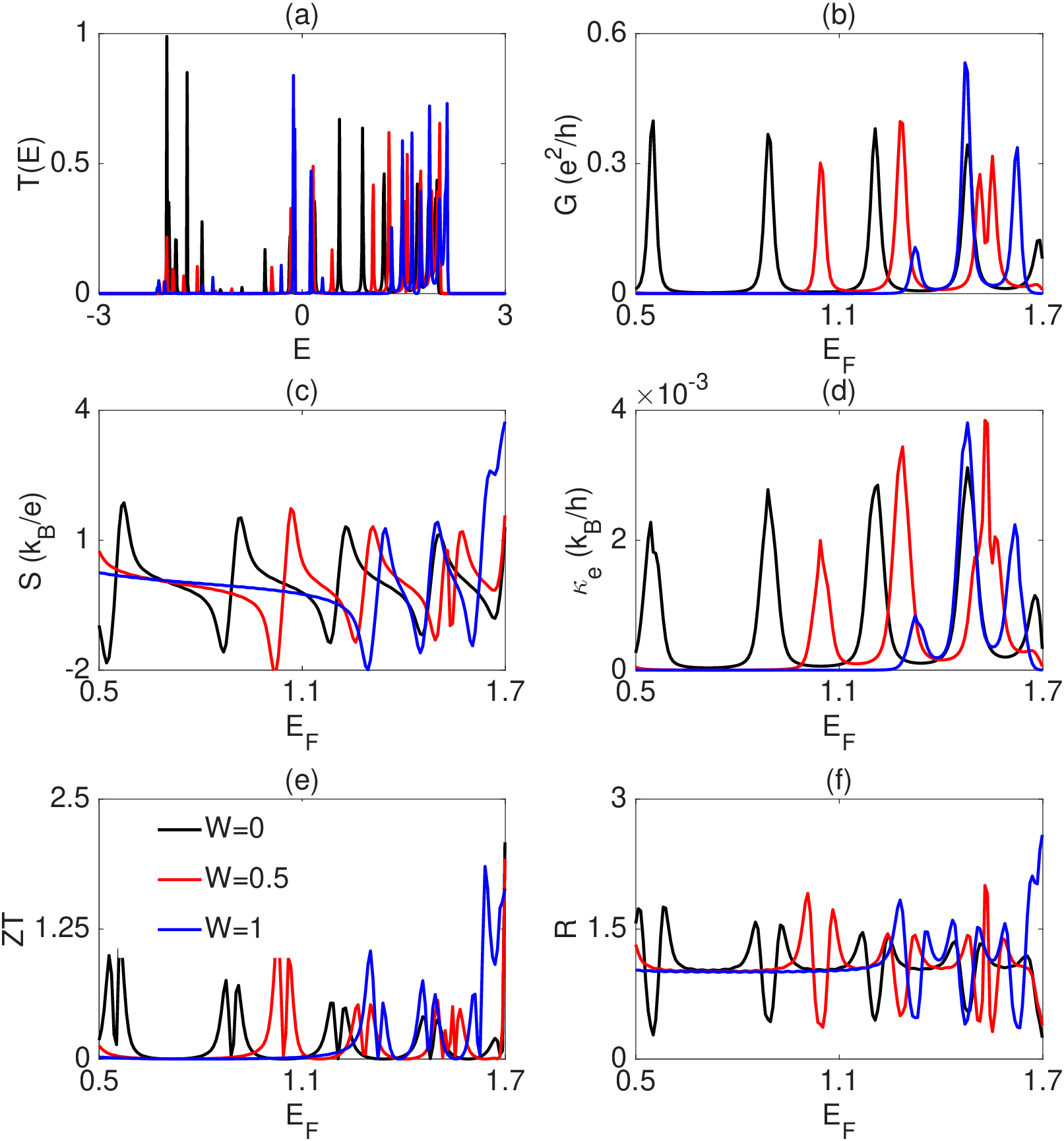}
    \caption{Asymmetric $(3-1)$ electrode-coupling geometry in the homogeneous-hopping regime ($t_1=t_2$). (a) Transmission spectrum, (b) electrical conductance, (c) Seebeck coefficient, (d) electronic thermal conductance, (e) thermoelectric figure of merit ($ZT$), and (f) Lorenz ratio as functions of the Fermi energy for different quasiperiodic modulation strengths. The parameters are $\gamma=0.05$, $T=0.005$, $\phi=0$, and $t_1=t_2=1$.}
    %\caption{Ring(3,1) $t_1$ Eq $t_2$}
    \label{fig7}
\end{figure}

\begin{figure}[t]
    \centering
    \includegraphics[width=8cm,height=8cm]{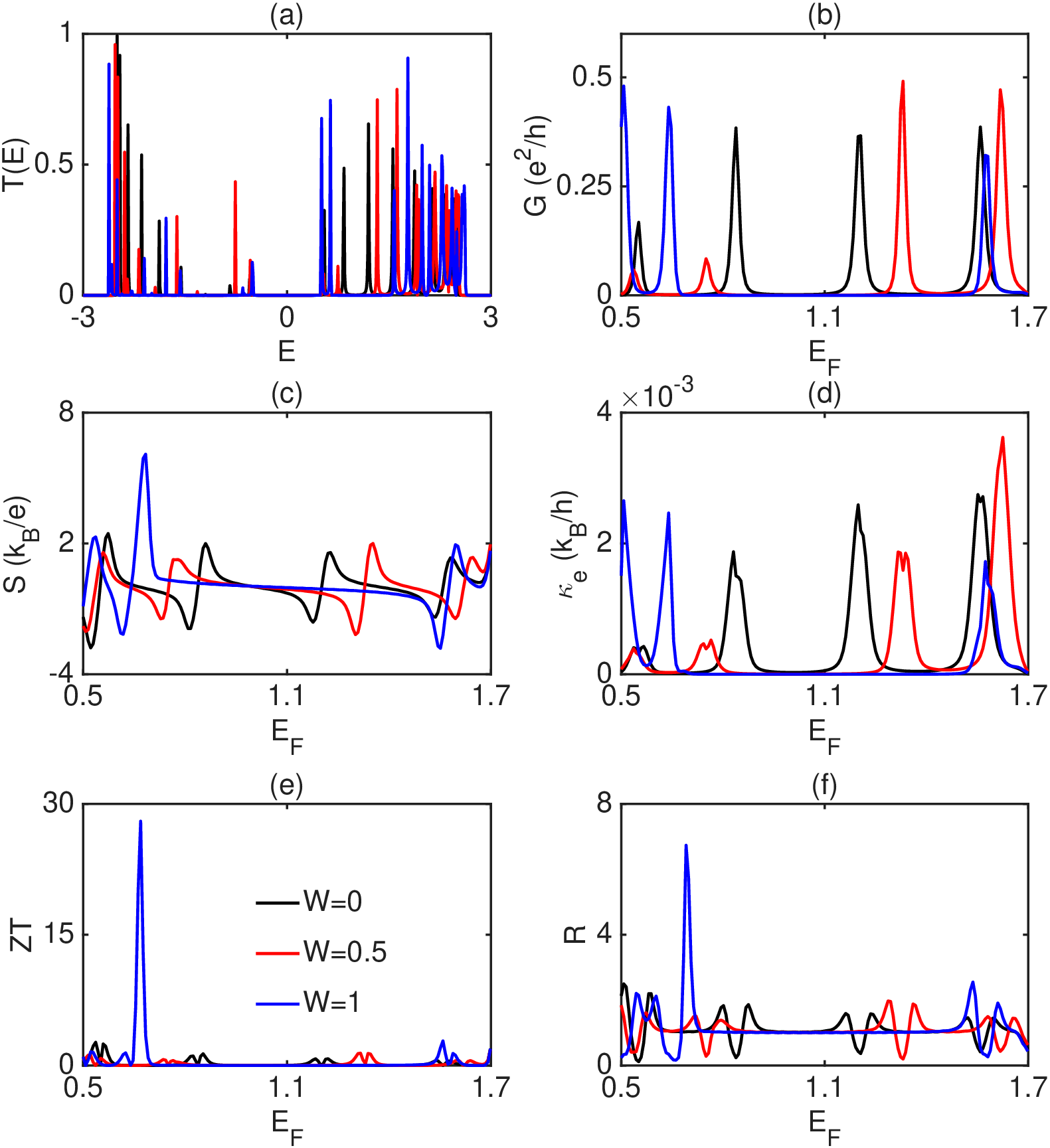}
    \caption{Asymmetric $(3-1)$ electrode-coupling geometry in the trivial dimerized regime ($t_1>t_2$). (a) Transmission spectrum, (b) electrical conductance, (c) Seebeck coefficient, (d) electronic thermal conductance, (e) thermoelectric figure of merit ($ZT$), and (f) Lorenz ratio as functions of the Fermi energy for different quasiperiodic modulation strengths. The parameters are $\gamma=0.05$, $T=0.005$, $\phi=0$, $t_1=1.5$, and $t_2=1$.}
    %\caption{Ring(3,1) $t_1$ gr $t_2$}
    \label{fig8}
\end{figure}

We now turn to the asymmetric electrode-coupling geometry, in which the source reservoir is
coupled to three lattice sites (1,2,N), whereas the drain is coupled to a single site (N/2+1). This
(3-1) configuration differs from the symmetric [3-3] geometry not in the underlying
quasiperiodic SSH ring, but solely in the spatial distribution of the system-electrode couplings.
At zero magnetic flux, $\phi=0$, this allows us to isolate the effect of coupling asymmetry on
coherent transport without the additional interference induced by the Aharonov-Bohm phase.
The reduced number of drain contacts restricts the number of coherent pathways contributing to
transmission and consequently modifies the interference pattern and energy selectivity of the
device. In what follows, we show that the interplay among coupling asymmetry,
quasiperiodicity, and hopping dimerization significantly enhances the thermoelectric response.

Figures \ref{fig6}-\ref{fig8} show the transmission function and the corresponding thermoelectric coefficients for the topological ($t_1<t_2$), homogeneous ($t_1=t_2$), and trivial dimerized ($t_1>t_2$) regimes, respectively, for different strengths of the quasiperiodic modulation $W$. A pronounced reduction in the number of transmission resonances is observed compared with the symmetric [3-3] geometry. Instead of a dense set of conducting channels, the asymmetric configuration yields a relatively small number of sharp, well-separated resonances. This reduction in the number of available pathways enhances the energy selectivity of transport and provides a more effective energy-filtering mechanism.

In the topological regime, transmission resonances [see Fig.~\ref{fig6}(a)] remain distributed over both positive and negative-energy regions, although their density is substantially reduced compared
with the symmetric geometry. The homogeneous regime exhibits still fewer and more weakly pronounced resonances [Fig.~\ref{fig7}(a)]. In the trivial dimerized regime, the transmission becomes even more selective, with a further reduction in the number and amplitude of the resonant channels [Fig.~\ref{fig8}(a)]. Thus, increasing dimerization in the $t_1>t_2$ direction, together with the
asymmetric contact geometry, progressively restricts the energy range over which efficient
electronic transmission occurs.

The electrical conductance follows the structure of the transmission spectrum, as expected from
the Landauer expression. The topological regime retains a relatively larger number of closely
spaced conductance peaks, whereas the homogeneous and trivial regimes exhibit fewer and
broader features [Figs.~\ref{fig6}(b), \ref{fig7}(b), and \ref{fig8}(b)]. Importantly, the reduction in electrical conductance is accompanied by a stronger suppression of the electronic thermal conductance. As the system evolves from the topological to the trivial dimerized regime, $\kappa_e$ is progressively reduced over a broad range of Fermi energies [Figs.~\ref{fig6}(d)-\ref{fig8}(d)]. This selective suppression of heat transport is particularly favorable for thermoelectric conversion because it reduces the electronic heat current without eliminating all charge transport.

The Seebeck coefficient provides a complementary measure of this enhanced energy selectivity.
Its magnitude is largest in the trivial dimerized regime, intermediate in the topological regime,
and smallest in the homogeneous system [Figs.~\ref{fig6}(c), \ref{fig7}(c), and \ref{fig8}(c)]. This behavior originates from the increasing energy asymmetry of the transmission function around the Fermi energy. Since the thermopower is determined by the first energy moment of the transmission within the thermal window, sharper and more asymmetric transmission features enhance the preferential
transport of carriers above or below the chemical potential. The combination of dimerization and
asymmetric electrode coupling, therefore, produces stronger energy filtering and consequently a
larger thermopower.

These trends are reflected directly in the thermoelectric figure of merit. The trivial dimerized
regime exhibits the strongest enhancement, reaching $ZT\simeq30$ for $W=1$, as shown in Fig.~\ref{fig8}(e). This large value results from the simultaneous enhancement of the Seebeck coefficient and suppression of the electronic thermal conductance. The topological regime gives an intermediate maximum of approximately $ZT\simeq8$, whereas the homogeneous regime reaches only $ZT\simeq2$. Thus, the thermoelectric enhancement is not determined solely by the magnitude of the electrical conductance; rather, it arises from the asymmetric coupling geometry's ability to generate strongly energy-selective transmission while preferentially suppressing heat-carrying electronic states.

A useful signature of this charge-heat decoupling is provided by the normalized Lorenz ratio
$R=L/L_0$. The strongest deviations from the Wiedemann-Franz law occur in the trivial dimerized
regime, while the topological regime also exhibits pronounced deviations. In contrast, the
homogeneous system remains comparatively closer to the conventional Wiedemann-Franz
behavior. Moreover, the regions of strong Lorenz-ratio suppression coincide with those in which
$ZT$ is enhanced [Figs.~\ref{fig6}(f)-\ref{fig8}(f)]. This correlation indicates that the asymmetric coupling geometry does not merely reduce the overall transmission; rather, it modifies the energy dependence of the transmission so that electrical and thermal transport are affected differently.

The overall effect is therefore a coupling-geometry-induced decoupling of charge and heat
transport. The [3-1] configuration reduces the number of coherent transmission pathways and
produces sharper, more selective resonances. When combined with SSH dimerization and
quasiperiodicity, this energy filtering enhances the thermopower while strongly suppressing
electronic heat transport. The resulting violation of the Wiedemann-Franz law provides a
microscopic signature of this decoupling and accompanies the enhancement of $ZT$. In particular,
the trivial dimerized regime benefits most strongly from this mechanism, reaching $ZT\sim30$ for
$W=1$, compared with the much smaller values obtained in the homogeneous regime.

Thus, even in the absence of magnetic flux, the spatial asymmetry of the electrode coupling provides an independent control parameter for coherent thermoelectric transport. By engineering the number and location of electrode contacts, one can reshape the transmission spectrum and selectively suppress heat-carrying channels, demonstrating that electrode geometry can serve as an effective route toward enhanced thermoelectric performance in quasiperiodic topological systems.

Table ~\ref{tab:ZT_comparison} summarizes the maximum thermoelectric figure of merit, $ZT_{\max}$, obtained for the different dimerization regimes and electrode-coupling geometries. The results demonstrate that both lattice dimerization and the spatial arrangement of the electrode contacts play important roles in determining the thermoelectric performance. In the symmetric [3-3] coupling geometry, multiple contacts at both electrodes provide multiple coherent transmission pathways, giving rise to a dense resonance structure and an enhanced thermoelectric response. Among the three
dimerization regimes, the trivial dimerized regime ($t_1>t_2$) exhibits the highest performance, with
$ZT_{\max}\simeq9$ at $W=1$. In contrast, the asymmetric [3-1] coupling geometry restricts the number of available transmission pathways by coupling the drain to a single lattice site, thereby yielding sharper, more isolated transmission resonances. This enhanced energy selectivity promotes stronger energy filtering and a more effective suppression of electronic thermal transport. Consequently, the trivial dimerized regime in the asymmetric geometry achieves a substantially larger $ZT_{\max} \simeq30$ at $W=1$. The comparison highlights that, although both coupling geometries favor the trivial dimerized phase at zero magnetic flux, the asymmetric [3-1] configuration provides a much stronger enhancement of thermoelectric performance through sharper energy filtering and enhanced charge-heat transport decoupling. These results demonstrate that electrode-coupling geometry, together with dimerization and quasiperiodicity, provides an effective route for engineering coherent thermoelectric transport in quasiperiodic SSH systems.

\begin{table*}[t]
\caption{Comparison of the maximum thermoelectric figure of merit ($ZT_{\rm max}$) for different hopping regimes and electrode-coupling geometries.}
\label{tab:ZT_comparison}
\begin{ruledtabular}
\begin{tabular}{cccc}
Hopping regime & Coupling geometry & Disorder strength &  $ZT_{max}$ \\
\hline
$t_1<t_2$ & $3$-$3$ symmetric & $W=1$ & $\sim 6$ \\
$t_1=t_2$ & $3$-$3$ symmetric & $W=1$ & $\sim 2$ \\
$t_1>t_2$ & $3$-$3$ symmetric & $W=1$ & $\sim 9$ \\
$t_1<t_2$ & $3$-$1$ asymmetric & $W=1$ & $\sim 8$ \\
$t_1=t_2$ & $3$-$1$ asymmetric & $W=1$ & $\sim 2$ \\
$t_1>t_2$ & $3$-$1$ asymmetric & $W=1$ & $\sim 30$ \\
\end{tabular}
\end{ruledtabular}
\end{table*}

\subsection{Symmetric electrode coupling: finite magnetic flux}

We next investigate the effect of a finite magnetic flux on coherent thermoelectric transport in
the quasiperiodic SSH ring with symmetric [3-3] electrode coupling. In this configuration, three
lattice sites are coupled to each of the source and drain reservoirs, providing multiple coherent
pathways through the ring. A magnetic flux $\Phi$ threading the ring introduces an Aharonov-Bohm
phase between these pathways and thereby provides an additional means of controlling their
constructive and destructive interference. Consequently, the transmission spectrum and the
associated thermoelectric coefficients become tunable by the applied flux. We consider the
topological ($t_1<t_2$), homogeneous ($t_1=t_2$), and dimerized ($t_1>t_2$) regimes, and characterize their response through the maximum electrical conductance, maximum Seebeck coefficient,
maximum electronic thermal conductance, maximum $ZT$, and maximum Lorenz ratio obtained by scanning the Fermi energy.

\begin{figure}[t]
    \centering
    \includegraphics[width=8cm,height=8cm]{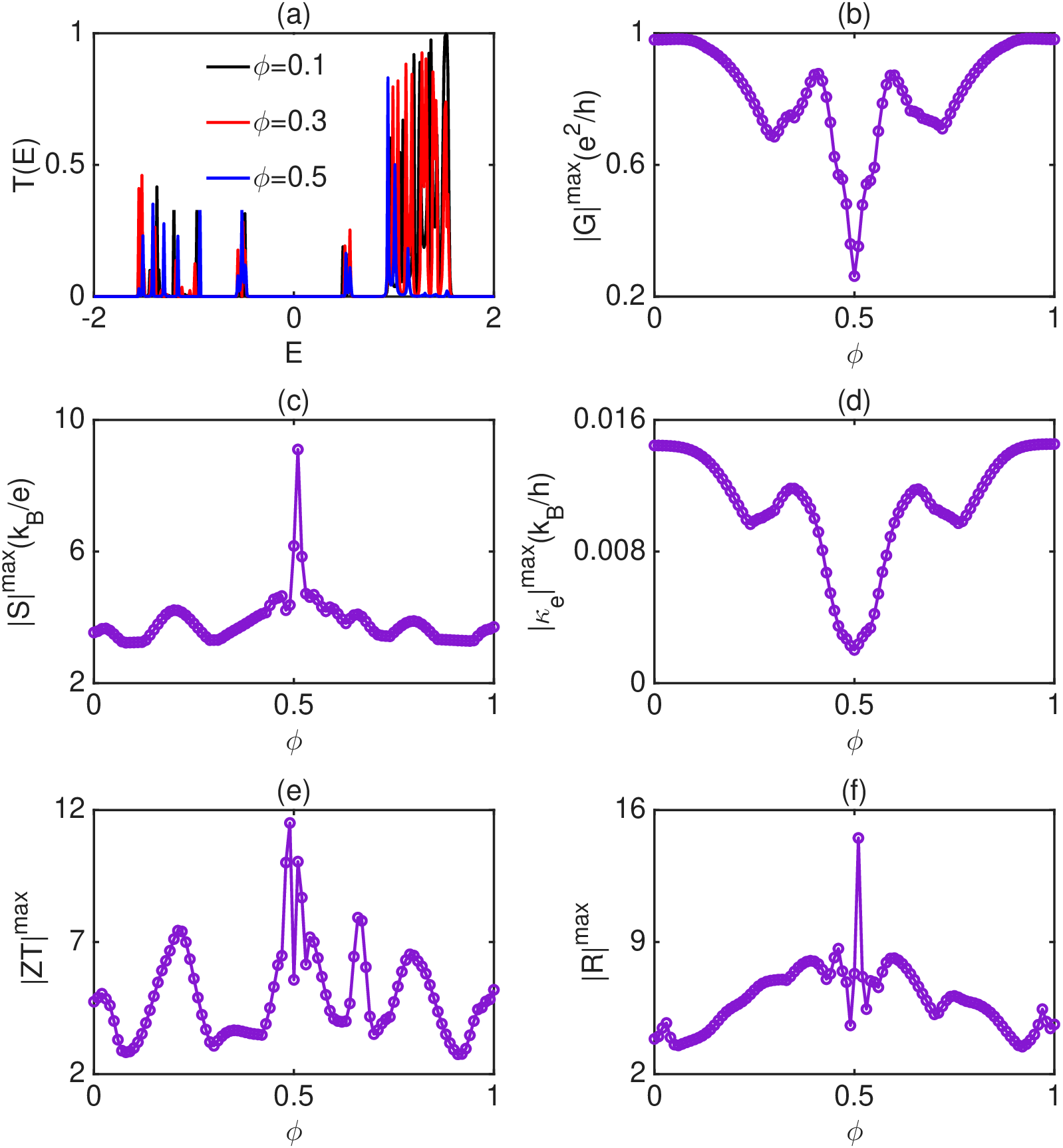}
    \caption{Symmetric electrode-coupling geometry in the topological dimerized regime ($t_1<t_2$). (a) Transmission spectrum for three representative magnetic flux values, $\phi=0.1$ (black), $\phi=0.3$ (red), and $\phi=0.5$ (blue). (b)–(f) Maximum electrical conductance, Seebeck coefficient, electronic thermal conductance, thermoelectric figure of merit ($ZT$), and Lorenz ratio, respectively, obtained by scanning the Fermi energy as functions of the magnetic flux. The parameters are $\gamma=0.05$, $T=0.005$, $t_1=0.5$, $t_2=1$, and $W=0.5$.}
    \label{fig9}
\end{figure}

\begin{figure}[t]
    \centering
    \includegraphics[width=8cm,height=8cm]{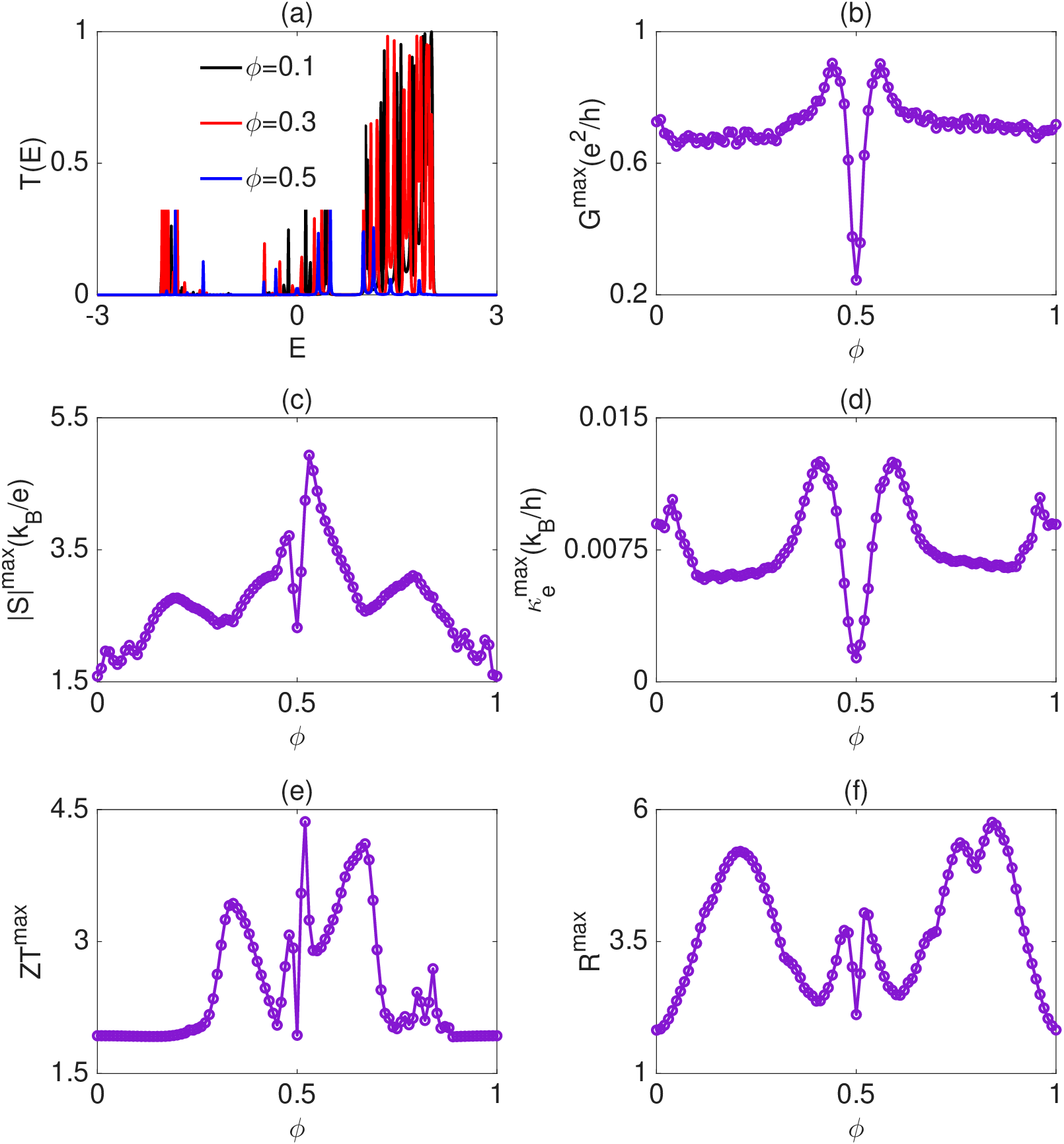}
    \caption{Symmetric electrode-coupling geometry in the homogeneous-hopping regime ($t_1=t_2$). (a) Transmission spectrum for three representative magnetic flux values, $\phi=0.1$ (black), $\phi=0.3$ (red), and $\phi=0.5$ (blue). (b)–(f) Maximum electrical conductance, Seebeck coefficient, electronic thermal conductance, thermoelectric figure of merit ($ZT$), and Lorenz ratio, respectively, obtained by scanning the Fermi energy as functions of the magnetic flux. The parameters are $\gamma=0.05$, $T=0.005$, $t_1=t_2=1$, and $W=0.5$.}
    %\caption{Ring(3,3) $t_1$ eq $t_2$}
    \label{fig10}
\end{figure}

\begin{figure}[t]
    \centering
    \includegraphics[width=8cm,height=8cm]{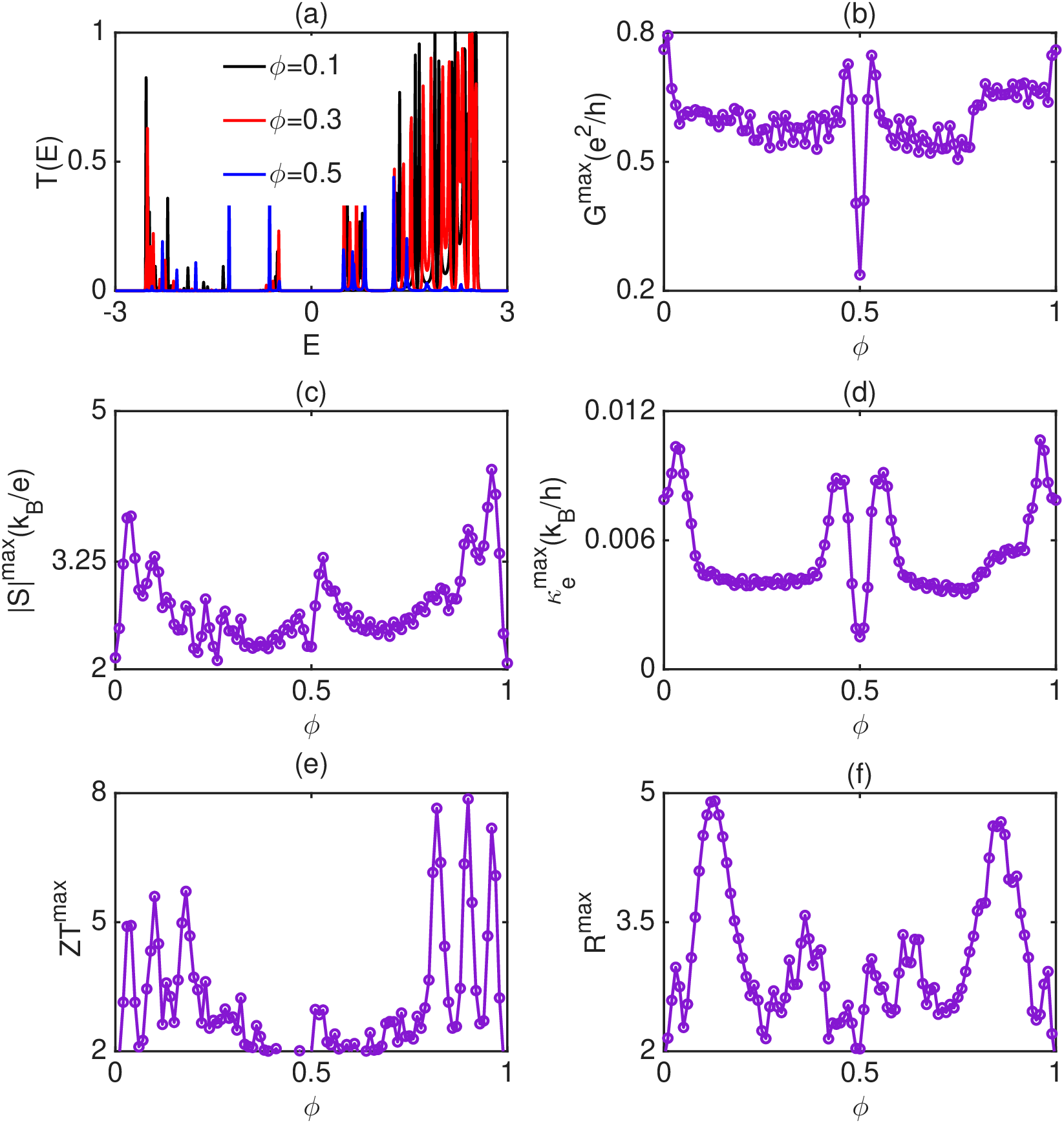}
    \caption{Symmetric electrode-coupling geometry in the trivial dimerized regime ($t_1>t_2$). (a) Transmission spectrum for three representative magnetic flux values, $\phi=0.1$ (black), $\phi=0.3$ (red), and $\phi=0.5$ (blue). (b)–(f) Maximum electrical conductance, Seebeck coefficient, electronic thermal conductance, thermoelectric figure of merit ($ZT$), and Lorenz ratio, respectively, obtained by scanning the Fermi energy as functions of the magnetic flux. The parameters are $\gamma=0.05$, $T=0.005$, $t_1=1.5$, $t_2=1$, and $W=0.5$.}
    %\caption{Ring(3,3) $t_1$ gr $t_2$}
    \label{fig11}
\end{figure}

The influence of magnetic flux on the coherent transmission is illustrated in Figs.~\ref{fig9}(a), \ref{fig10}(a), and \ref{fig11}(a) for the topological, homogeneous, and dimerized regimes, respectively. These panels show the transmission spectrum for representative flux values $\phi=0.1$, $0.3$, and $0.5$. In all three regimes, the magnetic flux substantially reconstructs the resonant transmission spectrum through Aharonov-Bohm interference. In the topological regime [Fig.~\ref{fig9}(a)], increasing the flux from $\phi=0.1$ to $0.3$ enhances both the number and amplitude of the transmission resonances, indicating predominantly constructive interference and improved coherent transport through the ring. Upon further increasing the flux toward $\phi=0.5$, the resonances are strongly suppressed, reflecting destructive interference between different propagation pathways.

A similar flux-induced reconstruction of the transmission spectrum occurs in the homogeneous
and trivial dimerized regimes, as shown in Figs.~\ref{fig10}(a) and ~\ref{fig11}(a), respectively. However, the detailed resonance structure depends strongly on the underlying hopping configuration. The dimerized systems retain a pronounced suppression of transmission around the band center,
reflecting the persistence of the dimerization-induced transport gap under moderate magnetic
flux. In contrast, the homogeneous system, lacking dimerization, exhibits transmission
resonances over a broader energy range. Thus, while the Aharonov-Bohm phase provides a
common mechanism for flux-dependent interference in all three cases, the resulting energy
landscape is strongly conditioned by the intrinsic dimerization of the SSH ring.

The flux-dependent reconstruction is particularly relevant for thermoelectric transport because
the transport coefficients are governed by the energy dependence of $T(E)$ around the Fermi
energy. The transmission spectrum in the topological regime exhibits a pronounced particle-hole
asymmetry, with a higher density of closely spaced resonances and larger transmission
amplitudes on the positive-energy side. These flux-tunable resonances provide enhanced energy
filtering and consequently influence the thermopower. The positive-energy region is therefore
particularly relevant for understanding the enhanced thermoelectric response.

The corresponding maximum electrical conductance is shown in Figs.~\ref{fig9}(b), ~\ref{fig10}(b), and ~\ref{fig11}(b). In all three hopping regimes, the conductance displays an approximately periodic dependence on the reduced magnetic flux $\phi=\Phi/\Phi_0$, with enhanced conductance near integer flux quanta and pronounced suppression near half-integer flux. This behavior is a characteristic consequence of Aharonov-Bohm interference: constructive interference near integer flux enhances coherent electron transmission, whereas destructive interference near half-integer flux suppresses the available conducting pathways.

The maximum electronic thermal conductance, shown in Figs.~\ref{fig9}(d),~\ref{fig10}(d), and ~\ref{fig11}(d), exhibits a qualitatively similar periodic dependence on magnetic flux, with a pronounced minimum around $\phi=0.5$. Its magnitude decreases systematically from the topological to the trivial dimerized regime. This behavior indicates that the combined action of lattice dimerization and magnetic-flux-induced interference can efficiently suppress heat transport, thereby providing favorable conditions for thermoelectric conversion.

The maximum absolute Seebeck coefficient, presented in Figs.~\ref{fig9}(c),~\ref{fig10}(c), and ~\ref{fig11}(c), shows a distinctly different dependence on the hopping regime. The thermopower is largest in the topological dimerized regime, intermediate in the homogeneous regime, and smallest in the trivial dimerized regime. This hierarchy originates from the flux-induced reconstruction of the
transmission spectrum and the associated energy asymmetry. The stronger asymmetry in the topological regime enhances energy filtering and yields a larger thermoelectric voltage, whereas the comparatively symmetric transmission in the homogeneous and trivial regimes results in a weaker thermopower.

The combined influence of charge and heat transport is reflected in the thermoelectric figure of
merit $ZT$, shown in Figs.~\ref{fig9}(e),~\ref{fig10}(e), and ~\ref{fig11}(e). The topological dimerized regime exhibits the strongest enhancement, reaching $ZT_{\max}\simeq 12$ near $\phi\simeq 0.48$. This enhancement arises from the simultaneous increase of the thermopower and suppression of electronic thermal conductance. The homogeneous regime reaches a substantially smaller $ZT_{\max}\simeq 4.5$, while the trivial dimerized regime displays an intermediate response. These results demonstrate that a large thermoelectric figure of merit requires an appropriate balance between energy-selective charge transport and suppression of heat conduction.

The maximum Lorenz ratio shown in Figs.~\ref{fig9}(f), ~\ref{fig10}(f), and ~\ref{fig11}(f) further reveals the breakdown of conventional charge-heat transport relations. Pronounced deviations from the Wiedemann-Franz law occur in both dimerized regimes, whereas the homogeneous regime remains comparatively closer to the universal Lorenz number. Notably, the strongest suppression of the Lorenz ratio occurs in the flux regions where $ZT$ is enhanced. This correlation demonstrates that magnetic-flux-induced quantum interference reshapes the energy dependence of the transmission in a
manner that affects electrical and thermal transport differently. The resulting charge-heat transport decoupling allows heat conduction to be suppressed more efficiently than charge
conduction, thereby enhancing the thermoelectric performance.

Overall, these results establish magnetic flux as an effective external control parameter for
coherent thermoelectric transport in the symmetric quasiperiodic SSH ring. By tuning the
Aharonov-Bohm phase, one can reconstruct the transmission spectrum, modify the coherent
transport pathways, and control the degree of energy filtering. Most importantly, the optimal
thermoelectric response shifts from the dimerized regime at zero flux, where $ZT_{\max}\simeq 9$, to the topological dimerized regime under finite flux, where $ZT_{\max}\simeq 12$ is obtained near half a flux quantum for the parameters considered in Fig.~\ref{fig9}. Thus, magnetic-flux-controlled quantum
interference provides an additional degree of freedom for optimizing the interplay between
charge and heat transport and for enhancing thermoelectric performance.

% \subsection{Interplay of magnetic flux and asymmetric electrode- coupling in thermoelectric transport}
\subsection{Asymmetric electrode coupling: Finite magnetic flux}
%\subsubsection{Dimerized regime: $t_1<t_2$}
\begin{figure}[t]
    \centering
    \includegraphics[width=8cm,height=8cm]{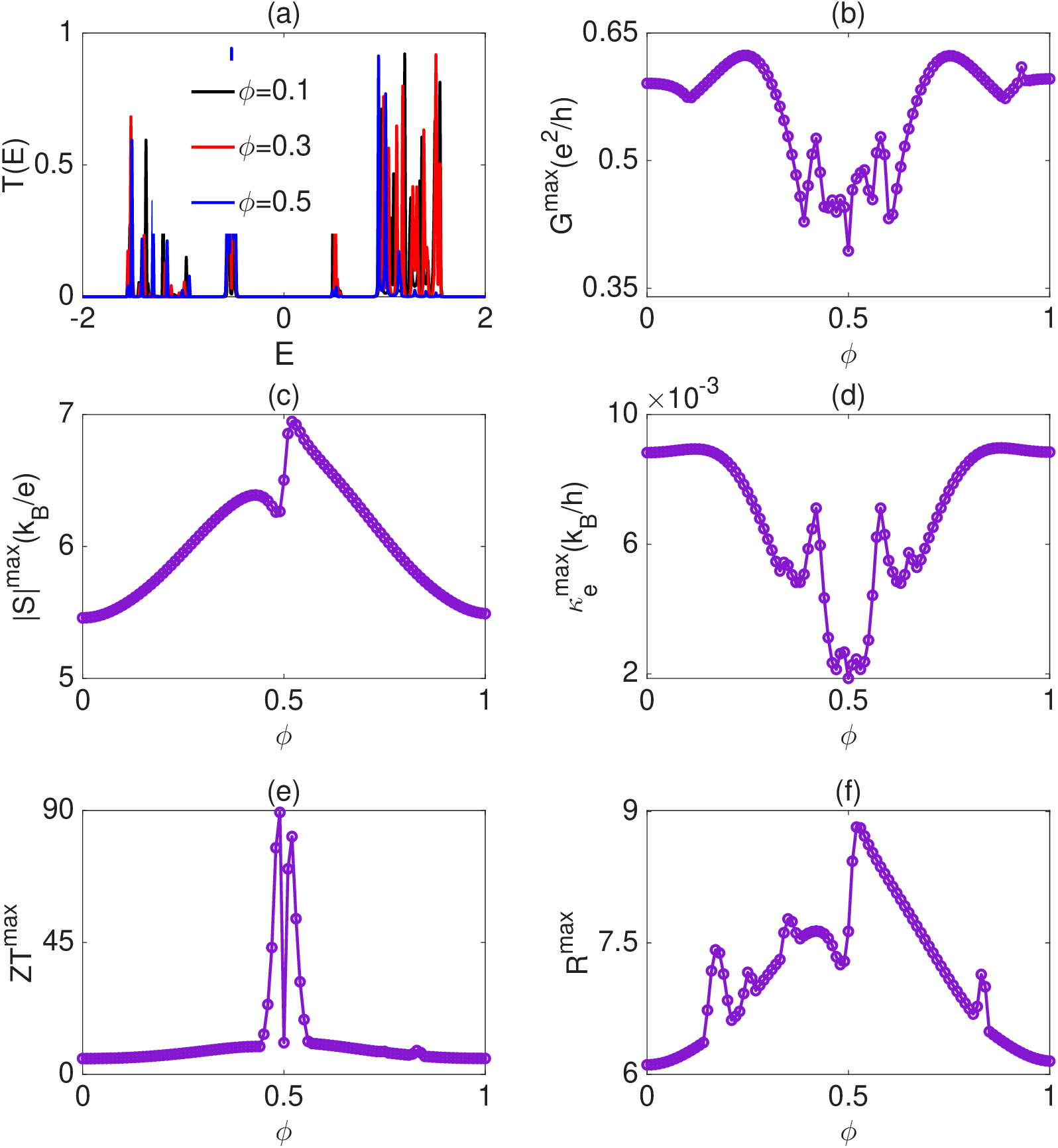}
    \caption{Asymmetric $(3,1)$ electrode-coupling geometry in the topological dimerized regime ($t_1<t_2$). (a) Transmission spectrum for three representative magnetic flux values, $\phi=0.1$ (black), $\phi=0.3$ (red), and $\phi=0.5$ (blue). (b)–(f) Maximum electrical conductance, Seebeck coefficient, electronic thermal conductance, thermoelectric figure of merit ($ZT$), and Lorenz ratio, respectively, obtained by scanning the Fermi energy as functions of the magnetic flux. The parameters are $\gamma=0.05$, $T=0.005$, $t_1=0.5$, $t_2=1$, and $W=0.5$.}
   % \caption{Ring(3,1) $t_1$ less $t_2$}
    \label{fig12}
\end{figure}
We finally consider the combined effect of magnetic flux and asymmetric electrode coupling on
coherent thermoelectric transport. In the asymmetric [3-1] configuration, three lattice sites are
coupled to the source reservoir, whereas only a single site is coupled to the drain. This broken
left-right coupling symmetry modifies the interference pathways available for electron
transmission and, in the presence of magnetic flux, provides a distinct mechanism for controlling
the Aharonov-Bohm interference. We examine the topological dimerized ($t_1<t_2$), homogeneous
($t_1=t_2$), and trivial dimerized ($t_1>t_2$) regimes by calculating the transmission spectrum and the corresponding maximum electrical conductance, Seebeck coefficient, electronic thermal
conductance, thermoelectric figure of merit, and Lorenz ratio as functions of the magnetic flux.

\begin{figure}[t]
    \centering
    \includegraphics[width=8cm,height=8cm]{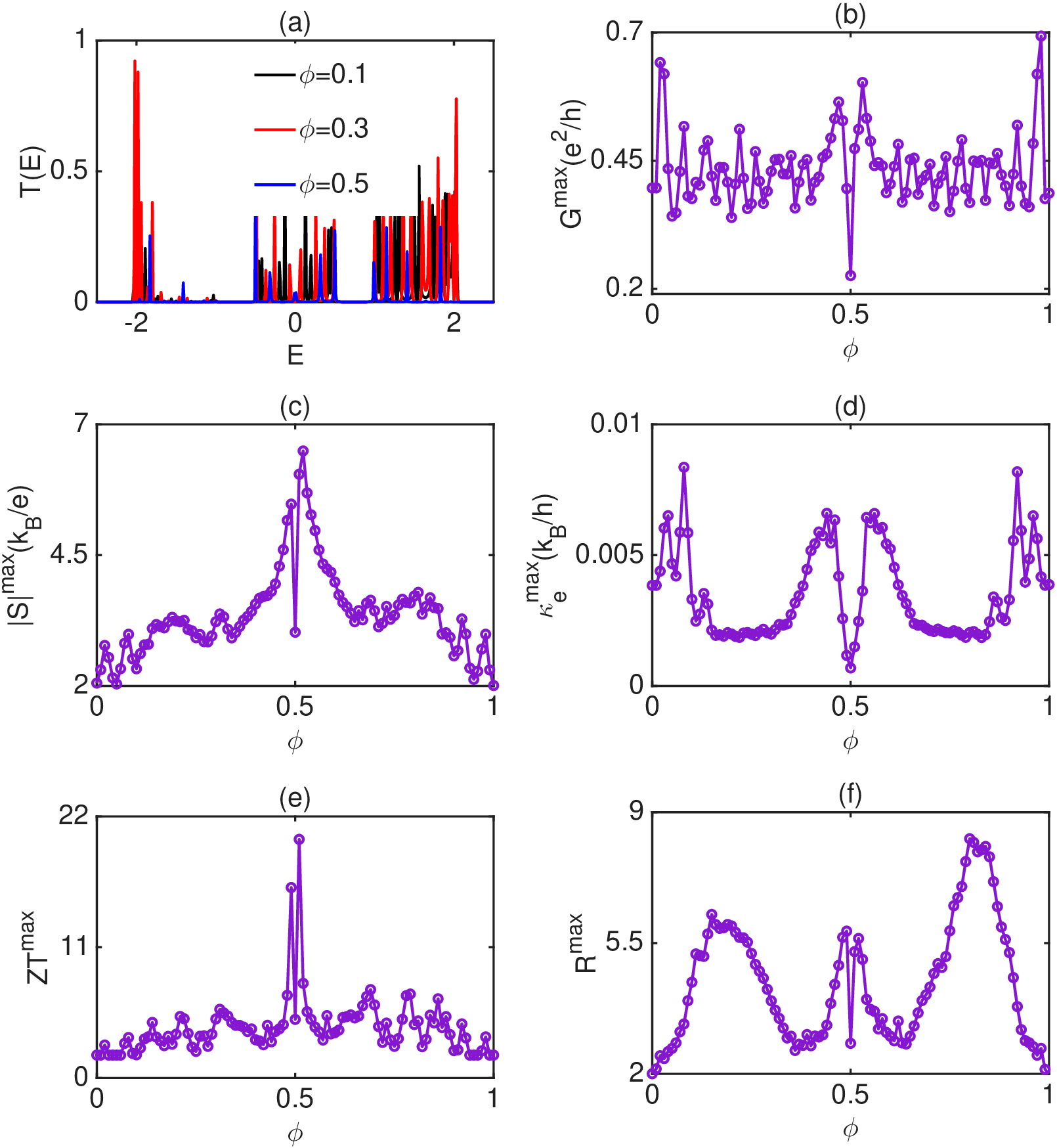}
    \caption{Asymmetric $(3,1)$ electrode-coupling geometry in the homogeneous-hopping regime ($t_1=t_2$). (a) Transmission spectrum for three representative magnetic flux values, $\phi=0.1$ (black), $\phi=0.3$ (red), and $\phi=0.5$ (blue). (b)–(f) Maximum electrical conductance, Seebeck coefficient, electronic thermal conductance, thermoelectric figure of merit ($ZT$), and Lorenz ratio, respectively, obtained by scanning the Fermi energy as functions of the magnetic flux. The parameters are $\gamma=0.05$, $T=0.005$, $t_1=t_2=1$, and $W=0.5$.}
   % \caption{Ring(3,1) $t_1$ Eq $t_2$}
    \label{fig13}
\end{figure}

\begin{figure}[t]
    \centering
    \includegraphics[width=8cm,height=8cm]{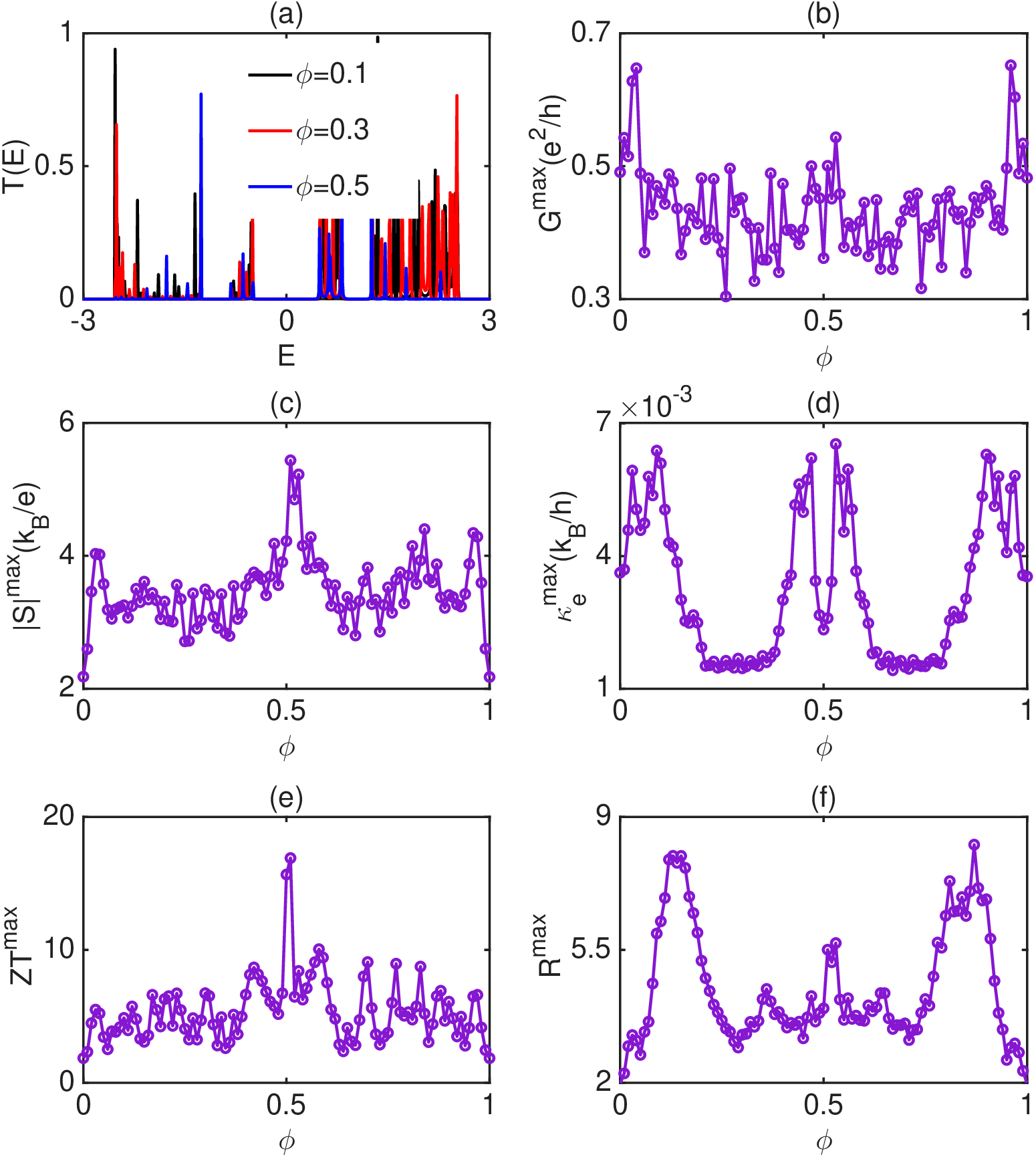}
    \caption{Asymmetric $(3,1)$ electrode-coupling geometry in the trivial dimerized regime ($t_1>t_2$). (a) Transmission spectrum for three representative magnetic flux values, $\phi=0.1$ (black), $\phi=0.3$ (red), and $\phi=0.5$ (blue). (b)–(f) Maximum electrical conductance, Seebeck coefficient, electronic thermal conductance, thermoelectric figure of merit ($ZT$), and Lorenz ratio, respectively, obtained by scanning the Fermi energy as functions of the magnetic flux. The parameters are $\gamma=0.05$, $T=0.005$, $t_1=1.5$, $t_2=1$, and $W=0.5$.}
    %\caption{Ring(3,1) $t_1$ gr $t_2$}
    \label{fig14}
\end{figure}

The flux-dependent transmission spectra for the three hopping regimes are shown in Figs.~\ref{fig12}(a),~\ref{fig13}(a), and ~\ref{fig14}(a) for the topological, homogeneous, and trivial dimerized regimes, respectively. In each case, the transmission is shown for representative flux values $\phi=0.1$, $0.3$, and $0.5$. The magnetic flux substantially reconstructs the resonance structure through Aharonov-Bohm interference, redistributing the available transmission channels. In the topological and trivial dimerized regimes, a pronounced suppression of transmission around the band center persists for all the flux values considered, reflecting the robustness of the dimerization-induced transport gap. By contrast, the homogeneous regime exhibits transmission resonances over a broader energy range due to the absence of a dimerization-induced gap.

The asymmetric coupling geometry leads to a particularly strong flux sensitivity of the
transmission spectrum. For the selected flux values, $\phi=0.3$ produces a comparatively large
number of transmission resonances, indicating enhanced constructive interference between the
available coherent pathways. As the flux approaches $\phi=0.5$, destructive Aharonov-Bohm
interference redistributes and suppresses the resonant channels. Because the thermoelectric
coefficients are determined by the energy dependence of the transmission around the Fermi
energy, this flux-controlled reconstruction provides an efficient mechanism for modifying the
energy filtering properties of the device.

The corresponding maximum electrical conductance is presented in Figs.~\ref{fig12}(b),~\ref{fig13}(b), and ~\ref{fig14}(b). In all three hopping regimes, the conductance exhibits a pronounced periodic dependence on the magnetic flux, reflecting the phase accumulated by electrons traversing the ring. The flux dependence is particularly sensitive to the underlying hopping configuration, with substantial oscillations arising from the redistribution of coherent transmission pathways. These results demonstrate that asymmetric electrode coupling does not eliminate the Aharonov-Bohm
response; rather, it modifies the interference pattern and consequently provides an additional
means of controlling coherent charge transport.

The maximum electronic thermal conductance, shown in Figs.~\ref{fig12}(d),~\ref{fig13}(d), and ~\ref{fig14}(d), decreases systematically from the topological to the trivial dimerized regime. The suppression of $\kappa_e$ reflects the combined influence of dimerization, asymmetric coupling, and magnetic-flux-induced interference on the energy-resolved transmission. In particular, reducing thermal transport while retaining appreciable electrical conduction is favorable for thermoelectric conversion. Thus, the asymmetric geometry provides a mechanism through which charge and heat currents can respond differently to the magnetic flux.

The maximum absolute Seebeck coefficient, shown in Figs.~\ref{fig12}(c),~\ref{fig13}(c), and~\ref{fig14}(c), is largest in the topological dimerized regime, followed by the homogeneous and trivial regimes. The enhanced thermopower in the topological phase originates from the pronounced energy asymmetry of its flux-dependent transmission spectrum. The broken left-right coupling
symmetry, together with the Aharonov-Bohm phase and topological dimerization, generates
stronger energy-selective transmission and consequently a larger thermoelectric voltage. In the
homogeneous and trivial regimes, the weaker transmission asymmetry results in comparatively
smaller values of the Seebeck coefficient.

The most striking consequence of this combined control is observed in the thermoelectric figure
of merit shown in Figs.~\ref{fig12}(e),~\ref{fig13}(e), and~\ref{fig14}(e). The topological dimerized regime exhibits a dramatic enhancement, reaching $ZT_{\max}\simeq90$ near $\phi\simeq0.48$ [see Table \ref{tab:2}]. This giant magnification of $ZT$ value results from the simultaneous enhancement of the Seebeck coefficient and strong suppression of the electronic thermal conductance, producing a favorable balance between charge and heat transport. By comparison, the homogeneous regime reaches $ZT_{\max}\simeq 20$, while the trivial dimerized regime reaches $ZT_{\max}\simeq 18$. Thus, in the presence of magnetic flux, the asymmetric (3-1) coupling geometry strongly favors the topological dimerized regime, in sharp contrast to the zero-flux case, where the trivial dimerized regime provides the strongest response.

The flux dependence of the Lorenz ratio, shown in Figs.~\ref{fig12}(f),~\ref{fig13}(f), and~\ref{fig14}(f), provides further evidence for the decoupling of charge and heat transport. Pronounced deviations from the Wiedemann-Franz law occur in all three hopping regimes, demonstrating that the magnetic flux and asymmetric coupling substantially modify the conventional relationship between electrical and thermal conductance. The strongest suppression of the Lorenz ratio occurs in the flux regions where $ZT$ is enhanced, establishing a direct correlation between the violation of the Wiedemann-Franz law and the thermoelectric performance. This behavior indicates that the

large $ZT$ is associated not simply with reduced transmission, but with a selective modification of
the transmission spectrum that suppresses heat transport more efficiently than charge transport.
A direct comparison with the symmetric geometry highlights the crucial role of electrode
asymmetry. In the topological dimerized regime, the symmetric [3-3] configuration reaches
$ZT_{\max}\simeq12$ at $\phi\simeq0.48$, whereas the asymmetric [3-1] configuration reaches $ZT_{\max}\simeq90$ at nearly the same flux. The enhancement arises from the different interference landscapes generated by the two coupling configurations. In the symmetric geometry, multiple pathways at both electrodes produce a relatively broad interference structure. In contrast, the asymmetric coupling breaks the left-right symmetry of the device and generates sharper energy-selective transmission features in the topological phase. These features enhance the Seebeck coefficient while strongly reducing the electronic thermal conductance, resulting in the giant thermoelectric response.

The overall results, therefore, establish electrode-coupling geometry and magnetic flux as
complementary control parameters for coherent thermoelectric transport. At zero magnetic flux,
the trivial dimerized regime provides the highest thermoelectric performance for both coupling
geometries, with $ZT_{\max}\simeq9$ for symmetric [3-3] coupling and $ZT_{\max}\simeq30$ for asymmetric [3-1] coupling at $W=1$. Under finite magnetic flux, however, Aharonov-Bohm interference changes the transmission landscape and shifts the optimal thermoelectric response to the topological
dimerized regime. In the asymmetric geometry, this effect is particularly pronounced, yielding
$ZT_{\max}\simeq90$ near $\phi\simeq0.48$, compared with approximately $ZT_{\max}\simeq12$ for the corresponding symmetric geometry for the same flux value.

Thus, the combination of topological dimerization, coupling asymmetry, and flux-controlled quantum interference provides a powerful route to engineering highly energy-selective transmission and to decoupling charge and heat transport. The resulting giant $ZT$ demonstrates that modifying the electrode geometry can amplify the thermoelectric consequences of Aharonov-Bohm interference far beyond those achievable with symmetric coupling alone.

%\subsubsection{Uniform hopping regime: $t_1=t_2$}

\begin{table*}[t]
\caption{Comparison of the maximum thermoelectric figure of merit ($ZT_{\rm max}$) for different hopping regimes and electrode-coupling geometries in the presence of magnetic flux at fixed $W=0.5$. Here, $\phi_{\rm max}$ denotes the magnetic flux at which $ZT_{\rm max}$ is attained.}
\label{tab:2}
\begin{ruledtabular}
\begin{tabular}{cccc}
Hopping regime & Coupling geometry & $\phi_{\rm max}$ & $ZT_{\rm max}$ \\
\hline
$t_1<t_2$ & Symmetric $(3,3)$  & $\sim0.48$ & $\sim12$ \\
$t_1=t_2$ & Symmetric $(3,3)$  & $\sim0.51$ & $\sim4.5$ \\
$t_1>t_2$ & Symmetric $(3,3)$  & $\sim0.46$ & $\sim8$ \\
$t_1<t_2$ & Asymmetric $(3,1)$ & $\sim0.48$ & $\sim90$ \\
$t_1=t_2$ & Asymmetric $(3,1)$ & $\sim0.52$ & $\sim20$ \\
$t_1>t_2$ & Asymmetric $(3,1)$ & $\sim0.52$ & $\sim18$ \\
\end{tabular}
\end{ruledtabular}
\end{table*}

\section{Conclusion}
\label{secIV}

In conclusion, we have demonstrated that coherent thermoelectric transport in a quasiperiodic
Su-Schrieffer-Heeger (SSH) ring can be profoundly reshaped by engineering the geometry of its
coupling to external reservoirs. Within a nonequilibrium Green's-function framework, we have
investigated the combined roles of SSH dimerization, quasiperiodic modulation, electrode-coupling asymmetry, and magnetic flux, and have shown that these ingredients provide complementary mechanisms for controlling the transmission spectrum and, consequently, the flow of charge and heat. Our results reveal that the electrode geometry is not merely a boundary condition for transport; it is an active element of quantum interference and can qualitatively alter the system's thermoelectric response.

At zero magnetic flux, the thermoelectric response is governed primarily by the interplay
between dimerization and the spatial distribution of the electrode couplings. The trivial
dimerized regime ($t_1>t_2$) provides the highest thermoelectric performance for both coupling
geometries. For the symmetric $[3-3]$ configuration, the maximum figure of merit is
$ZT_{\max}\simeq 9$, whereas asymmetric $[3-1]$ coupling enhances the energy selectivity of the
transmission and increases the figure of merit to approximately $ZT_{\max}\simeq 30$. This
enhancement demonstrates that reducing the number of available coherent pathways need not
simply suppress transport. Instead, appropriately engineered coupling asymmetry can sharpen the
relevant transmission features and thereby improve the balance between charge and heat
transport.

The introduction of magnetic flux reveals a qualitatively different aspect of this control. The
Aharonov-Bohm phase continuously modifies the relative phase accumulated along different
coherent pathways, reconstructing the transmission spectrum through constructive and
destructive interference. Consequently, the magnetic flux acts as an externally tunable spectral
filter. In the symmetric $[3-3]$ geometry, this flux-controlled interference shifts the optimal
thermoelectric response toward the topological dimerized regime ($t_1<t_2$), where
$ZT_{\max}\simeq 12$ is obtained near $\phi\simeq 0.48$. Thus, magnetic flux does more than
modulate the magnitude of transport: it changes which underlying electronic phase is most
favorable for thermoelectric conversion.

The most striking result emerges when magnetic flux is combined with asymmetric electrode
coupling. In the $[3-1]$ configuration, the topological dimerized regime becomes strongly
favored, yielding a giant thermoelectric figure of merit $ZT_{\max}\simeq 90$ near
$\phi\simeq 0.48$, compared with $ZT_{\max}\simeq 20$ and $\simeq 18$ in the homogeneous
and trivial regimes, respectively. The enhancement originates from a cooperative mechanism in
which topological dimerization, broken left-right coupling symmetry, and flux-induced quantum
interference jointly reshape the energy dependence of the transmission. The resulting energy
filtering simultaneously enhances the thermopower and suppresses the electronic thermal
conductance, producing a strong decoupling between charge and heat currents.

An important consequence of this mechanism is the pronounced violation of the Wiedemann-Franz law. The regions of strongest suppression of the Lorenz ratio correlate with those exhibiting enhanced $ZT$, indicating that the large thermoelectric response is intimately connected with the breakdown of the conventional proportionality between electrical and thermal transport. In this sense, the giant $ZT$ is not simply a consequence of high electrical conductance or low thermal conductance considered independently. Rather, it emerges from the selective manipulation of the transmission function, whereby electronic states that contribute efficiently to charge transport are retained while heat-carrying contributions are suppressed. This charge-heat decoupling provides the microscopic origin of the enhanced thermoelectric efficiency.

Our results, therefore, establish a useful hierarchy of control. Dimerization determines the
intrinsic electronic structure; the electrode coupling geometry determines the available coherent
pathways, and magnetic flux controls the relative phase between those pathways. Their combined action transforms the transmission function into a tunable energy filter and allows the optimal thermoelectric regime to be switched from the trivial dimerized phase at zero flux to the
topological dimerized phase under finite flux. The particularly large enhancement produced by
the asymmetric $[3-1]$ geometry demonstrates that the spatial engineering of reservoir coupling
can be as important as the intrinsic Hamiltonian parameters in designing quantum thermoelectric
devices.

More broadly, these findings suggest that quantum interference should be viewed not merely as a
source of oscillatory transport, but as a resource for engineering the spectral separation of charge
and heat currents. The combination of topology, quasiperiodicity, contact geometry, and Aharonov-Bohm interference offers a versatile platform in which thermoelectric performance can be controlled through both intrinsic and externally tunable parameters. The predicted effects should be relevant to experimentally accessible platforms such as gate-defined quantum-dot arrays, semiconductor nanowire quantum dots, molecular junctions, and cold-atom realizations of SSH-type systems, where hopping amplitudes, quasiperiodic potentials, electrode couplings, and magnetic flux can, in principle, be independently controlled.

The central message of this work is therefore that the geometry of quantum coupling can be elevated from a passive transport detail to an active design principle. By combining engineered electrode asymmetry with flux-controlled quantum interference and topological dimerization, one can dramatically reshape the energy-filtering properties of a mesoscopic system and achieve a giant thermoelectric response. The predicted $ZT_{\max}\simeq 90$ provides a particularly striking
manifestation of this principle and points toward a broader strategy for designing high-
performance of coherent thermoelectric devices through interference engineering rather than
material optimization alone.

\section*{ACKNOWLEDGMENTS}

Sridhar and Souvik Roy acknowledge financial support from the Indian Institute of Technology (IIT) Bhubaneswar through the Institute Research Fellowship.
\section*{DATA AVAILABILITY}
The data that support the findings of this study are available from the corresponding author upon reasonable request.

\bibliography{references}

@article{Mahan1996,
  author = {Mahan, G. D. and Sofo, J. O.},
  title = {The best thermoelectric},
  journal = {Proc. Natl. Acad. Sci. U.S.A.},
  volume = {93},
  pages = {7436--7439},
  year = {1996},
  doi = {10.1073/pnas.93.15.7436}
}

@article{Dresselhaus2007,
  author = {Dresselhaus, M. S. and Chen, G. and Tang, M. Y. and Yang, R. and Lee, H. and Wang, D. and Ren, Z. and Fleurial, J.-P. and Gogna, P.},
  title = {New directions for low-dimensional thermoelectric materials},
  journal = {Adv. Mater.},
  volume = {19},
  pages = {1043--1053},
  year = {2007},
  doi = {10.1002/adma.200600527}
}

@article{Hicks1993,
  author = {Hicks, L. D. and Dresselhaus, M. S.},
  title = {Effect of quantum-well structures on the thermoelectric figure of merit},
  journal = {Phys. Rev. B},
  volume = {47},
  pages = {12727--12731},
  year = {1993},
  doi = {10.1103/PhysRevB.47.12727}
}

@article{Karlstrom2011,
  author = {Karlstr\"om, O. and Linke, H. and Karlstr\"om, G. and Wacker, A.},
  title = {Increasing thermoelectric performance using coherent transport},
  journal = {Phys. Rev. B},
  volume = {84},
  pages = {113415},
  year = {2011},
  doi = {10.1103/PhysRevB.84.113415}
}

@article{Trocha2012,
  author = {Trocha, P. and Barna\'s, J.},
  title = {Large enhancement of thermoelectric effects in a double quantum dot system due to Fano resonance},
  journal = {Phys. Rev. B},
  volume = {85},
  pages = {085408},
  year = {2012},
  doi = {10.1103/PhysRevB.85.085408}
}

@book{Datta1995,
  author = {Datta, Supriyo},
  title = {Electronic Transport in Mesoscopic Systems},
  publisher = {Cambridge University Press},
  year = {1995}
}

@book{Datta2005,
  author = {Datta, Supriyo},
  title = {Quantum Transport: Atom to Transistor},
  publisher = {Cambridge University Press},
  year = {2005}
}

@book{Asboth2016,
  author    = {Asb{\'o}th, J{\'a}nos K. and Oroszl{\'a}ny, L{\'a}szl{\'o} and P{\'a}lyi, Andr{\'a}s},
  title     = {A Short Course on Topological Insulators},
  publisher = {Springer},
  year      = {2016},
  volume    = {919},
  series    = {Lecture Notes in Physics},
  doi       = {10.1007/978-3-319-25607-8}
}

@article{Harper1955,
  author = {Harper, P. G.},
  title = {Single band motion of conduction electrons in a uniform magnetic field},
  journal = {Proc. Phys. Soc. A},
  volume = {68},
  pages = {874},
  year = {1955},
  doi = {10.1088/0370-1298/68/10/304}
}

@article{svkprb2,
  title = {Non-Hermitian comb effect in coupled clean and quasiperiodic chains},
  author = {Padhi, Soumya Ranjan and Roy, Souvik and Paul, Biswajit and Banerjee, Sanchayan and Mishra, Tapan},
  journal = {Phys. Rev. B},
  volume = {114},
  issue = {2},
  pages = {024202},
  numpages = {11},
  year = {2026},
  month = {Jul},
  publisher = {American Physical Society},
  doi = {10.1103/5vmn-vsxf},
  url = {https://link.aps.org/doi/10.1103/5vmn-vsxf}
}

@article{Aharonov1959,
  author = {Aharonov, Y. and Bohm, D.},
  title = {Significance of electromagnetic potentials in quantum theory},
  journal = {Phys. Rev.},
  volume = {115},
  pages = {485--491},
  year = {1959},
  doi = {10.1103/PhysRev.115.485}
}

@article{Buttiker1984,
  author = {B\"uttiker, M. and Imry, Y. and Landauer, R.},
  title = {Josephson behavior in small normal one-dimensional rings},
  journal = {Phys. Lett. A},
  volume = {96},
  pages = {365--367},
  year = {1983},
  doi = {10.1016/0375-9601(83)90011-7}
}

@article{svkprb1,
  title = {Flux-driven charge and spin transport in a dimerized Hubbard ring with Fibonacci modulation},
  author = {Roy, Souvik and Padhi, Soumya Ranjan and Mishra, Tapan},
  journal = {Phys. Rev. B},
  volume = {113},
  issue = {15},
  pages = {155401},
  numpages = {11},
  year = {2026},
  month = {Apr},
  publisher = {American Physical Society},
  doi = {10.1103/9yh2-lkjp},
  url = {https://link.aps.org/doi/10.1103/9yh2-lkjp}
}

@article{Whitney2014,
  author = {Whitney, R. S.},
  title = {Most efficient quantum thermoelectric at finite power output},
  journal = {Phys. Rev. Lett.},
  volume = {112},
  pages = {130601},
  year = {2014},
  doi = {10.1103/PhysRevLett.112.130601}
}

@article{Benenti2017,
  author  = {Giuliano Benenti and Giulio Casati and Keiji Saito and Robert S. Whitney},
  title   = {Fundamental aspects of steady-state conversion of heat to work at the nanoscale},
  journal = {Physics Reports},
  volume  = {694},
  pages   = {1--124},
  year    = {2017},
  issn    = {0370-1573},
  doi     = {10.1016/j.physrep.2017.05.008}
}

@article{Bedkihal2025,
  author    = {Salil Bedkihal and Jayasmita Behera and Malay Bandyopadhyay},
  title     = {Fundamental aspects of Aharonov--Bohm quantum machines: thermoelectric heat engines and diodes},
  journal   = {Journal of Physics: Condensed Matter},
  volume    = {37},
  number    = {16},
  pages     = {163001},
  year      = {2025},
  doi       = {10.1088/1361-648X/adb921},
  publisher = {IOP Publishing}
}

@article{Bergfield2010,
  author    = {Justin P. Bergfield and Michelle A. Solis and Charles A. Stafford},
  title     = {Giant Thermoelectric Effect from Transmission Supernodes},
  journal   = {ACS Nano},
  year      = {2010},
  volume    = {4},
  number    = {9},
  pages     = {5314--5320},
  doi       = {10.1021/nn100490g},
  publisher = {American Chemical Society}
}

@article{Dhar2006,
  author    = {Abhishek Dhar and Diptiman Sen},
  title     = {Nonequilibrium Green's function formalism and the problem of bound states},
  journal   = {Physical Review B},
  volume    = {73},
  number    = {8},
  pages     = {085119},
  year      = {2006},
  publisher = {American Physical Society},
  doi       = {10.1103/PhysRevB.73.085119}
}

@article{Miao2018,
  author    = {Ruijiao Miao and Hailiang Xu and Maxim Skripnik and Longji Cui and Kun Wang and Kim G. L. Pedersen and Martin Leijnse and Fabian Pauly and Kenneth W{\"a}rnmark and Edgar Meyhofer and Pramod Reddy and Heiner Linke},
  title     = {Influence of Quantum Interference on the Thermoelectric Properties of Molecular Junctions},
  journal   = {Nano Letters},
  volume    = {18},
  number    = {9},
  pages     = {5666--5672},
  year      = {2018},
  doi       = {10.1021/acs.nanolett.8b02207},
  publisher = {American Chemical Society}
}

@article{GarciaSuarez2013,
  author    = {V. M. Garc{\'i}a-Su{\'a}rez and R. Ferrad{\'a}s and J. Ferrer},
  title     = {Impact of Fano and Breit-Wigner resonances in the thermoelectric properties of nanoscale junctions},
  journal   = {Physical Review B},
  volume    = {88},
  number    = {23},
  pages     = {235417},
  year      = {2013},
  publisher = {American Physical Society},
  doi       = {10.1103/PhysRevB.88.235417}
}

@article{Perroni2016,
  author    = {C. A. Perroni and D. Ninno and V. Cataudella},
  title     = {Thermoelectric efficiency of molecular junctions},
  journal   = {Journal of Physics: Condensed Matter},
  volume    = {28},
  number    = {37},
  pages     = {373001},
  year      = {2016},
  doi       = {10.1088/0953-8984/28/37/373001},
  publisher = {IOP Publishing}
}

@article{Lambert2015,
  author    = {Colin J. Lambert},
  title     = {Basic concepts of quantum interference and electron transport in single-molecule electronics},
  journal   = {Chemical Society Reviews},
  volume    = {44},
  number    = {4},
  pages     = {875--888},
  year      = {2015},
  publisher = {Royal Society of Chemistry},
  doi       = {10.1039/C4CS00203B}
}

@article{Miroshnichenko2010,
  author    = {A. E. Miroshnichenko and S. Flach and Y. S. Kivshar},
  title     = {Fano resonances in nanoscale structures},
  journal   = {Reviews of Modern Physics},
  volume    = {82},
  number    = {3},
  pages     = {2257--2298},
  year      = {2010},
  publisher = {American Physical Society},
  doi       = {10.1103/RevModPhys.82.2257}
}

@article{Bhattacharya2026Thermoelectric,
  author    = {Ranjini Bhattacharya and Souvik Roy},
  title     = {Thermoelectric enhancement via electronic and phononic channels in staggered and non-staggered dimerized quantum ring},
  journal   = {Physica E: Low-dimensional Systems and Nanostructures},
  volume    = {182},
  pages     = {116576},
  year      = {2026},
  month     = {July},
  doi       = {10.1016/j.physe.2026.116576},
  issn      = {1386-9477},
  publisher = {Elsevier}
}

@article{Roy2025FluxDriven,
  author    = {Souvik Roy and Santanu K. Maiti},
  title     = {Flux-Driven Circular Current in a Non-Hermitian Dimerized Aharonov--Bohm Ring: Impact of Physical Gain and Loss},
  journal   = {Annalen der Physik},
  volume    = {537},
  number    = {12},
  pages     = {e202500217},
  year      = {2025},
  doi       = {10.1002/andp.202500217},
  publisher = {Wiley}
}

@article{Roy2023CircularCurrent,
  author    = {Souvik Roy and Santanu K. Maiti},
  title     = {Circular current in a one-dimensional Hubbard quasi-periodic Su--Schrieffer--Heeger ring},
  journal   = {Journal of Physics: Condensed Matter},
  volume    = {35},
  number    = {35},
  pages     = {355303},
  year      = {2023},
  doi       = {10.1088/1361-648X/acd60f},
  publisher = {IOP Publishing}
}

@article{Chakraborty2024Thermoelectric,
  author    = {Suvendu Chakraborty and Santanu K. Maiti},
  title     = {Thermoelectric Phenomena in a Magnetic Heterostructure with AAH Modulation: Charge and Spin Figure of Merits},
  journal   = {Nanoscale and Microscale Thermophysical Engineering},
  volume    = {28},
  number    = {4},
  pages     = {129--146},
  year      = {2024},
  doi       = {10.1080/15567265.2024.2370839},
  publisher = {Taylor \& Francis}
}

@article{Ganguly2024Thermoelectric,
  author    = {Sudin Ganguly and Kallol Mondal and Santanu K. Maiti},
  title     = {Thermoelectric response in zigzag chains: Impact of irradiation-induced conformational changes},
  journal   = {Journal of Applied Physics},
  volume    = {136},
  number    = {1},
  pages     = {014301},
  year      = {2024},
  doi       = {10.1063/5.0213895},
  publisher = {AIP Publishing}
}

@article{Sridhar2026Interplay,
  author        = {Sridhar and Souvik Roy and Malay Bandyopadhyay},
  title         = {Interplay of Electrode Coupling Engineering, Quasiperiodicity, and Magnetic Flux in Quantum Transport through a Su--Schrieffer--Heeger Ring},
  journal       = {arXiv preprint},
  volume        = {arXiv:2606.28846},
  year          = {2026},
  eprint        = {2606.28846},
  archivePrefix = {arXiv},
  primaryClass  = {cond-mat.mes-hall},
  url           = {https://arxiv.org/abs/2606.28846}
}

@article{Bhandari2025Multidot,
  author    = {Tika Ram Bhandari and Hari Timsina and Chiranjibi Dhakal and Narayan Prasad Adhikari},
  title     = {Thermoelectric properties of the multi-dot quantum transport model},
  journal   = {Scientific Reports},
  volume    = {15},
  pages     = {21119},
  year      = {2025},
  doi       = {10.1038/s41598-025-05018-9},
  publisher = {Nature Publishing Group}
}

@article{Verma2022NonEquilibrium,
  author    = {Sachin Verma and Ajay Singh},
  title     = {Non-equilibrium thermoelectric transport across normal metal--Quantum dot--Superconductor hybrid system within the Coulomb blockade regime},
  journal   = {Journal of Physics: Condensed Matter},
  volume    = {34},
  number    = {15},
  pages     = {155601},
  year      = {2022},
  doi       = {10.1088/1361-648X/ac4ced},
  publisher = {IOP Publishing}
}

@article{Sitek2013Dicke,
  author    = {A. Sitek and A. Manolescu},
  title     = {Dicke states in multiple quantum dots},
  journal   = {Physical Review A},
  volume    = {88},
  number    = {4},
  pages     = {043807},
  year      = {2013},
  doi       = {10.1103/PhysRevA.88.043807},
  publisher = {American Physical Society}
}

@article{Bergfield2009Thermoelectric,
  author    = {J. P. Bergfield and C. A. Stafford},
  title     = {Thermoelectric Signatures of Coherent Transport in Single-Molecule Heterojunctions},
  journal   = {Nano Letters},
  volume    = {9},
  number    = {8},
  pages     = {3072--3076},
  year      = {2009},
  doi       = {10.1021/nl901554s},
  publisher = {American Chemical Society}
}

@article{Bennett2024Quantum,
  author    = {Runa X. Bennett and Joshua R. Hendrickson and Justin P. Bergfield},
  title     = {Quantum Interference Enhancement of the Spin-Dependent Thermoelectric Response},
  journal   = {ACS Nano},
  volume    = {18},
  number    = {18},
  pages     = {11876--11885},
  year      = {2024},
  doi       = {10.1021/acsnano.4c01297},
  publisher = {American Chemical Society}
}

@article{Behera2023Quantum,
  author    = {Jayasmita Behera and Salil Bedkihal and Bijay Kumar Agarwalla and Malay Bandyopadhyay},
  title     = {Quantum coherent control of nonlinear thermoelectric transport in a triple-dot Aharonov--Bohm heat engine},
  journal   = {Physical Review B},
  volume    = {108},
  number    = {16},
  pages     = {165419},
  year      = {2023},
  doi       = {10.1103/PhysRevB.108.165419},
  publisher = {American Physical Society}
}

@article{Bedkihal2013FluxDependent,
  author    = {Salil Bedkihal and Malay Bandyopadhyay and Dvira Segal},
  title     = {Flux-dependent occupations and occupation difference in geometrically symmetric and energy degenerate double-dot Aharonov--Bohm interferometers},
  journal   = {Physical Review B},
  volume    = {87},
  number    = {4},
  pages     = {045418},
  year      = {2013},
  doi       = {10.1103/PhysRevB.87.045418},
  publisher = {American Physical Society}
}

@article{Landauer1957,
  author    = {Rolf Landauer},
  title     = {Spatial Variation of Currents and Fields Due to Localized Scatterers in Metallic Conduction},
  journal   = {IBM Journal of Research and Development},
  volume    = {1},
  number    = {3},
  pages     = {223--231},
  year      = {1957},
  doi       = {10.1147/rd.13.0223}
}

@article{Buttiker1986,
  author    = {M. B{\"u}ttiker},
  title     = {Four-Terminal Phase-Coherent Conductance},
  journal   = {Physical Review Letters},
  volume    = {57},
  number    = {14},
  pages     = {1761--1764},
  year      = {1986},
  doi       = {10.1103/PhysRevLett.57.1761}
}

@article{Meir1992,
  author    = {Yigal Meir and Ned S. Wingreen},
  title     = {Landauer Formula for the Current through an Interacting Electron Region},
  journal   = {Physical Review Letters},
  volume    = {68},
  number    = {16},
  pages     = {2512--2515},
  year      = {1992},
  doi       = {10.1103/PhysRevLett.68.2512}
}

@book{Haug2008,
  author    = {Hartmut Haug and Antti-Pekka Jauho},
  title     = {Quantum Kinetics in Transport and Optics of Semiconductors},
  edition   = {2},
  publisher = {Springer},
  address   = {Berlin},
  year      = {2008},
  doi       = {10.1007/978-3-540-73564-9}
}

@article{Fisher1981,
  author    = {D. S. Fisher and P. A. Lee},
  title     = {Relation between Conductivity and Transmission Matrix},
  journal   = {Physical Review B},
  volume    = {23},
  number    = {12},
  pages     = {6851--6854},
  year      = {1981},
  doi       = {10.1103/PhysRevB.23.6851}
}

@article{Xu2014,
  author    = {Yong Xu and Zhongxue Gan and Shou-Cheng Zhang},
  title     = {Enhanced Thermoelectric Performance and Anomalous Seebeck Effects in Topological Insulators},
  journal   = {Physical Review Letters},
  volume    = {112},
  pages     = {226801},
  year      = {2014},
  doi       = {10.1103/PhysRevLett.112.226801}
}

@article{Zhou2025Interaction,
  author    = {Xiaofan Zhou and Suotang Jia and Jian-Song Pan},
  title     = {Interaction-induced phase transitions at topological quantum criticality of an extended Su--Schrieffer--Heeger model},
  journal   = {Physical Review B},
  volume    = {111},
  number    = {19},
  pages     = {195117},
  year      = {2025},
  doi       = {10.1103/PhysRevB.111.195117},
  publisher = {American Physical Society}
}

@article{Yamamoto2015,
  author    = {Kaoru Yamamoto and Naomichi Hatano},
  title     = {Thermodynamics of the mesoscopic thermoelectric heat engine beyond the linear-response regime},
  journal   = {Physical Review E},
  volume    = {92},
  number    = {4},
  pages     = {042165},
  year      = {2015},
  month     = {Oct},
  doi       = {10.1103/PhysRevE.92.042165},
  publisher = {American Physical Society}
}

@article{Sivan1986,
  author    = {Uri Sivan and Yoseph Imry},
  title     = {Multichannel Landauer formula for thermoelectric transport with application to thermopower near the mobility edge},
  journal   = {Physical Review B},
  volume    = {33},
  number    = {1},
  pages     = {551--558},
  year      = {1986},
  month     = {Jan},
  doi       = {10.1103/PhysRevB.33.551},
  publisher = {American Physical Society}
}

@article{Butcher1990,
  author    = {P. N. Butcher},
  title     = {Thermal and electrical transport formalism for electronic microstructures with many terminals},
  journal   = {Journal of Physics: Condensed Matter},
  volume    = {2},
  number    = {22},
  pages     = {4869--4878},
  year      = {1990},
  doi       = {10.1088/0953-8984/2/22/008},
  publisher = {IOP Publishing}
}

@article{Bedkihal2013,
  author    = {Salil Bedkihal and Malay Bandyopadhyay and Dvira Segal},
  title     = {The probe technique far from equilibrium: Magnetic field symmetries of nonlinear transport},
  journal   = {The European Physical Journal B},
  volume    = {86},
  number    = {12},
  pages     = {506},
  year      = {2013},
  doi       = {10.1140/epjb/e2013-40971-7},
  publisher = {Springer}
}

@article{PhysRevA.80.021603,
  title = {Localization in one-dimensional incommensurate lattices beyond the Aubry-Andr\'e model},
  author = {Biddle, J. and Wang, B. and Priour, D. J. and Das Sarma, S.},
  journal = {Phys. Rev. A},
  volume = {80},
  issue = {2},
  pages = {021603(R)},
  numpages = {4},
  year = {2009},
  month = {Aug},
  publisher = {American Physical Society},
  doi = {10.1103/PhysRevA.80.021603},
  url = {https://link.aps.org/doi/10.1103/PhysRevA.80.021603}
}

@article{l9gd-k9yw,
  title = {Coherent control of thermoelectric performance via engineered transmission functions in multidot Aharonov-Bohm heat engines},
  author = {Sridhar and Bedkihal, Salil and Bandyopadhyay, Malay},
  journal = {Phys. Rev. B},
  volume = {113},
  issue = {8},
  pages = {085428},
  numpages = {20},
  year = {2026},
  month = {Feb},
  publisher = {American Physical Society},
  doi = {10.1103/l9gd-k9yw},
  url = {https://link.aps.org/doi/10.1103/l9gd-k9yw}
}

@article{PhysRevLett.42.1698,
  title = {Solitons in Polyacetylene},
  author = {Su, W. P. and Schrieffer, J. R. and Heeger, A. J.},
  journal = {Phys. Rev. Lett.},
  volume = {42},
  issue = {25},
  pages = {1698--1701},
  numpages = {0},
  year = {1979},
  month = {Jun},
  publisher = {American Physical Society},
  doi = {10.1103/PhysRevLett.42.1698},
  url = {https://link.aps.org/doi/10.1103/PhysRevLett.42.1698}
}

@article{PhysRevB.22.2099,
  title = {Soliton excitations in polyacetylene},
  author = {Su, W. P. and Schrieffer, J. R. and Heeger, A. J.},
  journal = {Phys. Rev. B},
  volume = {22},
  issue = {4},
  pages = {2099--2111},
  numpages = {0},
  year = {1980},
  month = {Aug},
  publisher = {American Physical Society},
  doi = {10.1103/PhysRevB.22.2099},
  url = {https://link.aps.org/doi/10.1103/PhysRevB.22.2099}
}

@article{RevModPhys.60.781,
  title = {Solitons in conducting polymers},
  author = {Heeger, A. J. and Kivelson, S. and Schrieffer, J. R. and Su, W. -P.},
  journal = {Rev. Mod. Phys.},
  volume = {60},
  issue = {3},
  pages = {781--850},
  numpages = {0},
  year = {1988},
  month = {Jul},
  publisher = {American Physical Society},
  doi = {10.1103/RevModPhys.60.781},
  url = {https://link.aps.org/doi/10.1103/RevModPhys.60.781}
}

@article{PhysRevLett.112.130601,
  title = {Most Efficient Quantum Thermoelectric at Finite Power Output},
  author = {Whitney, Robert S.},
  journal = {Phys. Rev. Lett.},
  volume = {112},
  issue = {13},
  pages = {130601},
  numpages = {5},
  year = {2014},
  month = {Apr},
  publisher = {American Physical Society},
  doi = {10.1103/PhysRevLett.112.130601},
  url = {https://link.aps.org/doi/10.1103/PhysRevLett.112.130601}
}

@article{PhysRevB.91.115425,
  title = {Finding the quantum thermoelectric with maximal efficiency and minimal entropy production at given power output},
  author = {Whitney, Robert S.},
  journal = {Phys. Rev. B},
  volume = {91},
  issue = {11},
  pages = {115425},
  numpages = {18},
  year = {2015},
  month = {Mar},
  publisher = {American Physical Society},
  doi = {10.1103/PhysRevB.91.115425},
  url = {https://link.aps.org/doi/10.1103/PhysRevB.91.115425}
}

@article{PhysRevB.83.085428,
  title = {Optimal energy quanta to current conversion},
  author = {S\'anchez, Rafael and B\"uttiker, Markus},
  journal = {Phys. Rev. B},
  volume = {83},
  issue = {8},
  pages = {085428},
  numpages = {8},
  year = {2011},
  month = {Feb},
  publisher = {American Physical Society},
  doi = {10.1103/PhysRevB.83.085428},
  url = {https://link.aps.org/doi/10.1103/PhysRevB.83.085428}
}

@article{Sothmann_2015,
doi = {10.1088/0957-4484/26/3/032001},
url = {https://doi.org/10.1088/0957-4484/26/3/032001},
year = {2014},
month = {dec},
publisher = {IOP Publishing},
volume = {26},
number = {3},
pages = {032001},
author = {Sothmann, Björn and Sánchez, Rafael and Jordan, Andrew N},
title = {Thermoelectric energy harvesting with quantum dots},
journal = {Nanotechnology}
}

@article{PhysRevLett.106.230602,
  title = {Thermodynamic Bounds on Efficiency for Systems with Broken Time-Reversal Symmetry},
  author = {Benenti, Giuliano and Saito, Keiji and Casati, Giulio},
  journal = {Phys. Rev. Lett.},
  volume = {106},
  issue = {23},
  pages = {230602},
  numpages = {4},
  year = {2011},
  month = {Jun},
  publisher = {American Physical Society},
  doi = {10.1103/PhysRevLett.106.230602},
  url = {https://link.aps.org/doi/10.1103/PhysRevLett.106.230602}
}

\end{document}